\documentclass[longauth]{aa}  
\usepackage{url,hyperref}
\usepackage{footmisc}
\usepackage[normalem]{ulem}
\usepackage{graphicx}
\usepackage{units}
\usepackage{txfonts}
\usepackage{natbib}
\usepackage{ulem,color}

\usepackage[running,switch,columnwise]{lineno}

\bibpunct{(}{)}{;}{a}{}{,}

\renewcommand{\epsilon}{\varepsilon}

\newcommand{\vpsr}{PSR~B0833$-$45}
\newcommand{\hess}{H.E.S.S.}
\newcommand{\hessII}{\hess{}~II}
\newcommand{\fermi}{{\emph{Fermi}}}
\newcommand{\gr}{$\gamma$-ray}
\newcommand{\grs}{\gr{}s}
\makeatletter
\renewcommand*{\@fnsymbol}[1]{\ifcase#1\or*\or$\dagger$\or$\ddagger$\or**\or$\dagger\dagger$\or$\ddagger\ddagger$\fi}
\makeatother

\begin{document}

\title{First ground-based measurement of sub-20~GeV to 100~GeV \grs{} from the Vela pulsar with \hessII{}}

\author{\tiny{H.E.S.S. Collaboration
\and H.~Abdalla \inst{\ref{NWU}}
\and F.~Aharonian \inst{\ref{MPIK},\ref{DIAS},\ref{NASRA}}
\and F.~Ait~Benkhali \inst{\ref{MPIK}}
\and E.O.~Ang\"uner \inst{\ref{CPPM}}
\and M.~Arakawa \inst{\ref{Rikkyo}}
\and C.~Arcaro \inst{\ref{NWU}}
\and C.~Armand \inst{\ref{LAPP}}
\and M.~Arrieta \inst{\ref{LUTH}}
\and M.~Backes \inst{\ref{UNAM},\ref{NWU}}
\and M.~Barnard \inst{\ref{NWU}}
\and Y.~Becherini \inst{\ref{Linnaeus}}
\and J.~Becker~Tjus \inst{\ref{RUB}}
\and D.~Berge \inst{\ref{DESY}}
\and S.~Bernhard \inst{\ref{LFUI}}
\and K.~Bernl\"ohr \inst{\ref{MPIK}}
\and R.~Blackwell \inst{\ref{Adelaide}}
\and M.~B\"ottcher \inst{\ref{NWU}}
\and C.~Boisson \inst{\ref{LUTH}}
\and J.~Bolmont \inst{\ref{LPNHE}}
\and S.~Bonnefoy \inst{\ref{DESY}}
\and P.~Bordas \inst{\ref{MPIK}}
\and J.~Bregeon \inst{\ref{LUPM}}
\and F.~Brun \inst{\ref{CENB}}
\and P.~Brun \inst{\ref{IRFU}}
\and M.~Bryan \inst{\ref{GRAPPA}}
\and M.~B\"{u}chele \inst{\ref{ECAP}}
\and T.~Bulik \inst{\ref{UWarsaw}}
\and T.~Bylund \inst{\ref{Linnaeus}}
\and M.~Capasso \inst{\ref{IAAT}}
\and S.~Caroff \inst{\ref{LLR}}
\and A.~Carosi \inst{\ref{LAPP}}
\and S.~Casanova \inst{\ref{IFJPAN},\ref{MPIK}}
\and M.~Cerruti \inst{\ref{LPNHE}}
\and N.~Chakraborty \inst{\ref{MPIK}}
\and S.~Chandra \inst{\ref{NWU}}
\and R.C.G.~Chaves \inst{\ref{LUPM},\ref{CurieChaves}}
\and A.~Chen \inst{\ref{WITS}}
\and S.~Colafrancesco \inst{\ref{WITS}}
\and B.~Condon \inst{\ref{CENB}}
\and I.D.~Davids \inst{\ref{UNAM}}
\and C.~Deil \inst{\ref{MPIK}}
\and J.~Devin \inst{\ref{LUPM}}
\and P.~deWilt \inst{\ref{Adelaide}}
\and L.~Dirson \inst{\ref{HH}}
\and A.~Djannati-Ata\"i \inst{\ref{APC}}\protect\footnotemark[1]
\and A.~Dmytriiev \inst{\ref{LUTH}}
\and A.~Donath \inst{\ref{MPIK}}
\and V.~Doroshenko \inst{\ref{IAAT}}
\and L.O'C.~Drury \inst{\ref{DIAS}}
\and J.~Dyks \inst{\ref{NCAC}}
\and K.~Egberts \inst{\ref{UP}}
\and G.~Emery \inst{\ref{LPNHE}}
\and J.-P.~Ernenwein \inst{\ref{CPPM}}
\and S.~Eschbach \inst{\ref{ECAP}}
\and S.~Fegan \inst{\ref{LLR}}
\and A.~Fiasson \inst{\ref{LAPP}}
\and G.~Fontaine \inst{\ref{LLR}}
\and S.~Funk \inst{\ref{ECAP}}
\and M.~F\"u{\ss}ling \inst{\ref{DESY}}
\and S.~Gabici \inst{\ref{APC}}
\and Y.A.~Gallant \inst{\ref{LUPM}}
\and F.~Gat{\'e} \inst{\ref{LAPP}}
\and G.~Giavitto \inst{\ref{DESY}}\protect\footnotemark[1]
\and D.~Glawion \inst{\ref{LSW}}
\and J.F.~Glicenstein \inst{\ref{IRFU}}
\and D.~Gottschall \inst{\ref{IAAT}}
\and M.-H.~Grondin \inst{\ref{CENB}}
\and J.~Hahn \inst{\ref{MPIK}}
\and M.~Haupt \inst{\ref{DESY}}
\and G.~Heinzelmann \inst{\ref{HH}}
\and G.~Henri \inst{\ref{Grenoble}}
\and G.~Hermann \inst{\ref{MPIK}}
\and J.A.~Hinton \inst{\ref{MPIK}}
\and W.~Hofmann \inst{\ref{MPIK}}
\and C.~Hoischen \inst{\ref{UP}}
\and T.~L.~Holch \inst{\ref{HUB}}
\and M.~Holler \inst{\ref{LFUI}}\protect\footnotemark[1]
\and D.~Horns \inst{\ref{HH}}
\and D.~Huber \inst{\ref{LFUI}}
\and H.~Iwasaki \inst{\ref{Rikkyo}}
\and A.~Jacholkowska \inst{\ref{LPNHE}} \protect\footnotemark[2] 
\and M.~Jamrozy \inst{\ref{UJK}}
\and D.~Jankowsky \inst{\ref{ECAP}}
\and F.~Jankowsky \inst{\ref{LSW}}
\and L.~Jouvin \inst{\ref{APC}}
\and I.~Jung-Richardt \inst{\ref{ECAP}}
\and M.A.~Kastendieck \inst{\ref{HH}}
\and K.~Katarzy{\'n}ski \inst{\ref{NCUT}}
\and M.~Katsuragawa \inst{\ref{KAVLI}}
\and U.~Katz \inst{\ref{ECAP}}
\and D.~Kerszberg \inst{\ref{LPNHE}}
\and D.~Khangulyan \inst{\ref{Rikkyo}}
\and B.~Kh\'elifi \inst{\ref{APC}}
\and J.~King \inst{\ref{MPIK}}
\and S.~Klepser \inst{\ref{DESY}}
\and W.~Klu\'{z}niak \inst{\ref{NCAC}}
\and Nu.~Komin \inst{\ref{WITS}}
\and K.~Kosack \inst{\ref{IRFU}}
\and S.~Krakau \inst{\ref{RUB}}
\and M.~Kraus \inst{\ref{ECAP}}
\and P.P.~Kr\"uger \inst{\ref{NWU}}
\and G.~Lamanna \inst{\ref{LAPP}}
\and J.~Lau \inst{\ref{Adelaide}}
\and J.~Lefaucheur \inst{\ref{IRFU}}
\and A.~Lemi\`ere \inst{\ref{APC}}
\and M.~Lemoine-Goumard \inst{\ref{CENB}}
\and J.-P.~Lenain \inst{\ref{LPNHE}}
\and E.~Leser \inst{\ref{UP}}
\and T.~Lohse \inst{\ref{HUB}}
\and M.~Lorentz \inst{\ref{IRFU}}
\and R.~L\'opez-Coto \inst{\ref{MPIK}}
\and I.~Lypova \inst{\ref{DESY}}
\and D.~Malyshev \inst{\ref{IAAT}}
\and V.~Marandon \inst{\ref{MPIK}}
\and A.~Marcowith \inst{\ref{LUPM}}
\and C.~Mariaud \inst{\ref{LLR}}
\and G.~Mart\'i-Devesa \inst{\ref{LFUI}}
\and R.~Marx \inst{\ref{MPIK}}
\and G.~Maurin \inst{\ref{LAPP}}
\and P.J.~Meintjes \inst{\ref{UFS}}
\and A.M.W.~Mitchell \inst{\ref{MPIK}}
\and R.~Moderski \inst{\ref{NCAC}}
\and M.~Mohamed \inst{\ref{LSW}}
\and L.~Mohrmann \inst{\ref{ECAP}}
\and E.~Moulin \inst{\ref{IRFU}}
\and T.~Murach \inst{\ref{DESY}}
\and S.~Nakashima  \inst{\ref{RIKKEN}}
\and M.~de~Naurois \inst{\ref{LLR}}
\and H.~Ndiyavala  \inst{\ref{NWU}}
\and F.~Niederwanger \inst{\ref{LFUI}}
\and J.~Niemiec \inst{\ref{IFJPAN}}
\and L.~Oakes \inst{\ref{HUB}}
\and P.~O'Brien \inst{\ref{Leicester}}
\and H.~Odaka \inst{\ref{Tokyo}}
\and S.~Ohm \inst{\ref{DESY}}
\and M.~Ostrowski \inst{\ref{UJK}}
\and I.~Oya \inst{\ref{DESY}}
\and M.~Padovani \inst{\ref{LUPM}}
\and M.~Panter \inst{\ref{MPIK}}
\and R.D.~Parsons \inst{\ref{MPIK}}
\and C.~Perennes \inst{\ref{LPNHE}}
\and P.-O.~Petrucci \inst{\ref{Grenoble}}
\and B.~Peyaud \inst{\ref{IRFU}}
\and Q.~Piel \inst{\ref{LAPP}}
\and S.~Pita \inst{\ref{APC}}
\and V.~Poireau \inst{\ref{LAPP}}
\and A.~Priyana~Noel \inst{\ref{UJK}}
\and D.A.~Prokhorov \inst{\ref{WITS}}
\and H.~Prokoph \inst{\ref{DESY}}
\and G.~P\"uhlhofer \inst{\ref{IAAT}}
\and M.~Punch \inst{\ref{APC},\ref{Linnaeus}}
\and A.~Quirrenbach \inst{\ref{LSW}}
\and S.~Raab \inst{\ref{ECAP}}
\and R.~Rauth \inst{\ref{LFUI}}
\and A.~Reimer \inst{\ref{LFUI}}
\and O.~Reimer \inst{\ref{LFUI}}
\and M.~Renaud \inst{\ref{LUPM}}
\and F.~Rieger \inst{\ref{MPIK}}
\and L.~Rinchiuso \inst{\ref{IRFU}}
\and C.~Romoli \inst{\ref{MPIK}}
\and G.~Rowell \inst{\ref{Adelaide}}
\and B.~Rudak \inst{\ref{NCAC}}
\and E.~Ruiz-Velasco \inst{\ref{MPIK}}
\and V.~Sahakian \inst{\ref{YPI},\ref{NASRA}}
\and S.~Saito \inst{\ref{Rikkyo}}
\and D.A.~Sanchez \inst{\ref{LAPP}}
\and A.~Santangelo \inst{\ref{IAAT}}
\and M.~Sasaki \inst{\ref{ECAP}}
\and R.~Schlickeiser \inst{\ref{RUB}}
\and F.~Sch\"ussler \inst{\ref{IRFU}}
\and A.~Schulz \inst{\ref{DESY}}
\and U.~Schwanke \inst{\ref{HUB}}
\and S.~Schwemmer \inst{\ref{LSW}}
\and M.~Seglar-Arroyo \inst{\ref{IRFU}}
\and M.~Senniappan \inst{\ref{Linnaeus}}
\and A.S.~Seyffert \inst{\ref{NWU}}
\and N.~Shafi \inst{\ref{WITS}}
\and I.~Shilon \inst{\ref{ECAP}}
\and K.~Shiningayamwe \inst{\ref{UNAM}}
\and R.~Simoni \inst{\ref{GRAPPA}}
\and A.~Sinha \inst{\ref{APC}}
\and H.~Sol \inst{\ref{LUTH}}
\and F.~Spanier \inst{\ref{NWU}}
\and A.~Specovius \inst{\ref{ECAP}}
\and M.~Spir-Jacob \inst{\ref{APC}}
\and {\L.}~Stawarz \inst{\ref{UJK}}
\and R.~Steenkamp \inst{\ref{UNAM}}
\and C.~Stegmann \inst{\ref{UP},\ref{DESY}}
\and C.~Steppa \inst{\ref{UP}}
\and T.~Takahashi  \inst{\ref{KAVLI}}
\and J.-P.~Tavernet \inst{\ref{LPNHE}}
\and T.~Tavernier \inst{\ref{IRFU}}\protect\footnotemark[1]
\and A.M.~Taylor \inst{\ref{DESY}}
\and R.~Terrier \inst{\ref{APC}}
\and L.~Tibaldo \inst{\ref{MPIK}}
\and D.~Tiziani \inst{\ref{ECAP}}
\and M.~Tluczykont \inst{\ref{HH}}
\and C.~Trichard \inst{\ref{LLR}}
\and M.~Tsirou \inst{\ref{LUPM}}
\and N.~Tsuji \inst{\ref{Rikkyo}}
\and R.~Tuffs \inst{\ref{MPIK}}
\and Y.~Uchiyama \inst{\ref{Rikkyo}}
\and D.J.~van~der~Walt \inst{\ref{NWU}}
\and C.~van~Eldik \inst{\ref{ECAP}}
\and C.~van~Rensburg \inst{\ref{NWU}}
\and B.~van~Soelen \inst{\ref{UFS}}
\and G.~Vasileiadis \inst{\ref{LUPM}}
\and J.~Veh \inst{\ref{ECAP}}
\and C.~Venter \inst{\ref{NWU}}\protect\footnotemark[1]
\and P.~Vincent \inst{\ref{LPNHE}}
\and J.~Vink \inst{\ref{GRAPPA}}
\and F.~Voisin \inst{\ref{Adelaide}}
\and H.J.~V\"olk \inst{\ref{MPIK}}
\and T.~Vuillaume \inst{\ref{LAPP}}
\and Z.~Wadiasingh \inst{\ref{NWU}}
\and S.J.~Wagner \inst{\ref{LSW}}
\and R.M.~Wagner \inst{\ref{OKC}}
\and R.~White \inst{\ref{MPIK}}
\and A.~Wierzcholska \inst{\ref{IFJPAN}}
\and R.~Yang \inst{\ref{MPIK}}
\and D.~Zaborov \inst{\ref{LLR}}
\and M.~Zacharias \inst{\ref{NWU}}
\and R.~Zanin \inst{\ref{MPIK}}
\and A.A.~Zdziarski \inst{\ref{NCAC}}
\and A.~Zech \inst{\ref{LUTH}}
\and F.~Zefi \inst{\ref{LLR}}
\and A.~Ziegler \inst{\ref{ECAP}}
\and J.~Zorn \inst{\ref{MPIK}}
\and N.~\.Zywucka \inst{\ref{UJK}}
\\
and
\\
M.~Kerr \inst{\ref{NRL}}, S. Johnston \inst{\ref{CSIRO}} \and R.M.~Shannon \inst{\ref{CSIRO},\ref{OZGRAV}} 
}
}

\institute{
Centre for Space Research, North-West University, Potchefstroom 2520, South Africa \label{NWU} \and 
Universit\"at Hamburg, Institut f\"ur Experimentalphysik, Luruper Chaussee 149, D 22761 Hamburg, Germany \label{HH} \and 
Max-Planck-Institut f\"ur Kernphysik, P.O. Box 103980, D 69029 Heidelberg, Germany \label{MPIK} \and 
Dublin Institute for Advanced Studies, 31 Fitzwilliam Place, Dublin 2, Ireland \label{DIAS} \and 
National Academy of Sciences of the Republic of Armenia,  Marshall Baghramian Avenue, 24, 0019 Yerevan, Republic of Armenia  \label{NASRA} \and
Yerevan Physics Institute, 2 Alikhanian Brothers St., 375036 Yerevan, Armenia \label{YPI} \and
Institut f\"ur Physik, Humboldt-Universit\"at zu Berlin, Newtonstr. 15, D 12489 Berlin, Germany \label{HUB} \and
University of Namibia, Department of Physics, Private Bag 13301, Windhoek, Namibia \label{UNAM} \and
GRAPPA, Anton Pannekoek Institute for Astronomy, University of Amsterdam,  Science Park 904, 1098 XH Amsterdam, The Netherlands \label{GRAPPA} \and
Department of Physics and Electrical Engineering, Linnaeus University,  351 95 V\"axj\"o, Sweden \label{Linnaeus} \and
Institut f\"ur Theoretische Physik, Lehrstuhl IV: Weltraum und Astrophysik, Ruhr-Universit\"at Bochum, D 44780 Bochum, Germany \label{RUB} \and
Institut f\"ur Astro- und Teilchenphysik, Leopold-Franzens-Universit\"at Innsbruck, A-6020 Innsbruck, Austria \label{LFUI} \and
School of Physical Sciences, University of Adelaide, Adelaide 5005, Australia \label{Adelaide} \and
LUTH, Observatoire de Paris, PSL Research University, CNRS, Universit\'e Paris Diderot, 5 Place Jules Janssen, 92190 Meudon, France \label{LUTH} \and
Sorbonne Universit\'e, Universit\'e Paris Diderot, Sorbonne Paris Cit\'e, CNRS/IN2P3, Laboratoire de Physique Nucl\'eaire et de Hautes Energies, LPNHE, 4 Place Jussieu, F-75252 Paris, France \label{LPNHE} \and
Laboratoire Univers et Particules de Montpellier, Universit\'e Montpellier, CNRS/IN2P3,  CC 72, Place Eug\`ene Bataillon, F-34095 Montpellier Cedex 5, France \label{LUPM} \and
IRFU, CEA, Universit\'e Paris-Saclay, F-91191 Gif-sur-Yvette, France \label{IRFU} \and
Astronomical Observatory, The University of Warsaw, Al. Ujazdowskie 4, 00-478 Warsaw, Poland \label{UWarsaw} \and
Aix Marseille Universit\'e, CNRS/IN2P3, CPPM, Marseille, France \label{CPPM} \and
Instytut Fizyki J\c{a}drowej PAN, ul. Radzikowskiego 152, 31-342 Krak{\'o}w, Poland \label{IFJPAN} \and
Funded by EU FP7 Marie Curie, grant agreement No. PIEF-GA-2012-332350 \label{CurieChaves}  \and
School of Physics, University of the Witwatersrand, 1 Jan Smuts Avenue, Braamfontein, Johannesburg, 2050 South Africa \label{WITS} \and
Laboratoire d'Annecy de Physique des Particules, Univ. Grenoble Alpes, Univ. Savoie Mont Blanc, CNRS, LAPP, 74000 Annecy, France \label{LAPP} \and
Landessternwarte, Universit\"at Heidelberg, K\"onigstuhl, D 69117 Heidelberg, Germany \label{LSW} \and
Universit\'e Bordeaux, CNRS/IN2P3, Centre d'\'Etudes Nucl\'eaires de Bordeaux Gradignan, 33175 Gradignan, France \label{CENB} \and
Oskar Klein Centre, Department of Physics, Stockholm University, Albanova University Center, SE-10691 Stockholm, Sweden \label{OKC} \and
Institut f\"ur Astronomie und Astrophysik, Universit\"at T\"ubingen, Sand 1, D 72076 T\"ubingen, Germany \label{IAAT} \and
Laboratoire Leprince-Ringuet, Ecole Polytechnique, CNRS/IN2P3, F-91128 Palaiseau, France \label{LLR} \and
APC, AstroParticule et Cosmologie, Universit\'{e} Paris Diderot, CNRS/IN2P3, CEA/Irfu, Observatoire de Paris, Sorbonne Paris Cit\'{e}, 10, rue Alice Domon et L\'{e}onie Duquet, 75205 Paris Cedex 13, France \label{APC} \and
Univ. Grenoble Alpes, CNRS, IPAG, F-38000 Grenoble, France \label{Grenoble} \and
Department of Physics and Astronomy, The University of Leicester, University Road, Leicester, LE1 7RH, United Kingdom \label{Leicester} \and
Nicolaus Copernicus Astronomical Center, Polish Academy of Sciences, ul. Bartycka 18, 00-716 Warsaw, Poland \label{NCAC} \and
Iinstitut f\"ur Physik und Astronomie, Universit\"at Potsdam,  Karl-Liebknecht-Strasse 24/25, D 14476 Potsdam, Germany \label{UP} \and
Friedrich-Alexander-Universit\"at Erlangen-N\"urnberg, Erlangen Centre for Astroparticle Physics, Erwin-Rommel-Str. 1, D 91058 Erlangen, Germany \label{ECAP} \and
DESY, D-15738 Zeuthen, Germany \label{DESY} \and
Obserwatorium Astronomiczne, Uniwersytet Jagiello{\'n}ski, ul. Orla 171, 30-244 Krak{\'o}w, Poland \label{UJK} \and
Centre for Astronomy, Faculty of Physics, Astronomy and Informatics, Nicolaus Copernicus University,  Grudziadzka 5, 87-100 Torun, Poland \label{NCUT} \and
Department of Physics, University of the Free State,  PO Box 339, Bloemfontein 9300, South Africa \label{UFS} \and
Department of Physics, Rikkyo University, 3-34-1 Nishi-Ikebukuro, Toshima-ku, Tokyo 171-8501, Japan \label{Rikkyo} \and
Kavli Institute for the Physics and Mathematics of the Universe (Kavli IPMU), The University of Tokyo Institutes for Advanced Study (UTIAS), The University of Tokyo, 5-1-5 Kashiwa-no-Ha, Kashiwa City, Chiba, 277-8583, Japan \label{KAVLI} \and
Department of Physics, The University of Tokyo, 7-3-1 Hongo, Bunkyo-ku, Tokyo 113-0033, Japan \label{Tokyo} \and
RIKEN, 2-1 Hirosawa, Wako, Saitama 351-0198, Japan \label{RIKKEN} \and
Now at The School of Physics, The University of New South Wales, Sydney, 2052, Australia \label{MaxtedNowAt} \and
Now at Instituto de F\'{i}sica de S\~{a}o Carlos, Universidade de S\~{a}o Paulo, Av. Trabalhador S\~{a}o-carlense, 400 - CEP 13566-590, S\~{a}o Carlos, SP, Brazil \label{VianaNowAt} \and
Space Science Division, Naval Research Laboratory, Washington, DC 20375-5352, USA \label{NRL} \and
CSIRO Astronomy and Space Science, Australia Telescope National Facility, PO~Box~76, Epping NSW~1710, Australia \label{CSIRO} \and
Centre for Astrophysics and Supercomputing, Swinburne University of
Technology, Mail H30, PO Box 218, Hawthorn, VIC 3122, Australia.
ARC Centre of Excellence for Gravitational Wave Discovery (OzGrav)\label{OZGRAV}
}

\offprints{H.E.S.S.~collaboration,
\protect\\\email{\href{mailto:contact.hess@hess-experiment.eu}{contact.hess@hess-experiment.eu}};
\protect\\\protect\footnotemark[1] Corresponding authors
\protect\\\protect\footnotemark[2] Deceased
}

\date{Version 1.4, 15.02.2018}
\abstract
{}
{We report on the measurement and investigation of pulsed high-energy $\gamma$-ray emission from the Vela pulsar, \vpsr, based on observations  
with the largest telescope of \hess,  CT5, in monoscopic mode, and on data obtained with the \fermi-LAT.}
{Data from 40.3 hours of observations carried out with the \hessII{} array from 2013 to 2015 have been used.  
A dedicated very low-threshold event reconstruction and analysis pipeline was developed to achieve the lowest possible energy threshold. 
Eight years of \fermi-LAT data were analysed and also used as reference to validate the CT5 telescope response model and analysis methods.}
{A pulsed $\gamma$-ray signal at a significance level of more than $15\sigma$ is detected from the P2 peak of the Vela pulsar light curve.
Of a total of 15835 events, more than 6000 lie at an energy below 20~GeV, implying a significant overlap between \hessII{}-CT5 and the \fermi-LAT.
While the investigation of the pulsar light curve with the LAT confirms characteristics previously known up to 20~GeV
in the tens of GeV energy range, CT5 data show a change in the pulse morphology of P2, i.e. an extreme sharpening of its trailing edge, together with the possible onset of a new component at 3.4$\sigma$ significance level.
Assuming a power-law model for the P2 spectrum, an excellent agreement is found for the photon indices ($\Gamma \simeq$ 4.1) 
obtained with the two telescopes above 10~GeV and   
an upper bound of 8\% is derived on the relative offset between their energy scales.
Using data from both instruments, it is shown however that the spectrum of P2 in the 10-100~GeV has a pronounced curvature; this is a confirmation of the sub-exponential cut-off form found at lower energies with the LAT. This is further supported by weak evidence of an emission above 100~GeV obtained with CT5. 
In contrast, converging indications are found from both CT5 and LAT data for the emergence of a hard component 
above 50~GeV in the leading wing (LW2) of P2, which possibly extends beyond 100~GeV.}
{The detection demonstrates the performance and understanding of CT5 from 100~GeV down to the sub-20~GeV domain, i.e. unprecedented low energy for ground-based $\gamma$-ray astronomy. 
The extreme sharpening of the trailing edge of the P2 peak found in the \hessII{} light curve of the Vela pulsar and the possible extension beyond 100~GeV of 
at least one of its features, LW2, provide further constraints to models of \gr\  emission from pulsars.}

\authorrunning{H.E.S.S. Collaboration}
\titlerunning{Sub-20~GeV to 100~GeV pulsations from the Vela pulsar with \hessII{}}
\maketitle
\makeatletter
\renewcommand*{\@fnsymbol}[1]{\ifcase#1\@arabic{#1}\fi}
\makeatother
%
\section{Introduction}
The Vela pulsar, \vpsr{}, was one of the very first \gr{} sources
discovered with the SAS-II mission \citep{Thompson1975} and  has since been detected with subsequent space-borne \gr{} instruments, namely, 
{\it COS B} \citep{VelaCosBKanbach80}, {\it EGRET} \citep{VelaEgretKanbach94}, {\it  AGILE} \citep{AgilePulsars2009}, and \fermi-LAT \citep{fermifirstvela}.
At a period of 89~ms, the light curve of the pulsar exhibits two peaks, labelled P1 and P2, separated by 0.43 in phase and connected by a bridge emission, labelled P3. 

The initial detection of the Vela pulsar with the \fermi-LAT instrument was based on 75 days of verification and early phase observations and reached 
energies above $\unit[10]{GeV}$ \citep{fermifirstvela}. With 11 months of data, a high significance signal was obtained up to $\unit[20]{GeV}$ \citep{Abdo2010Vela}.   
More recently, \citet{Leung14} exploiting a deeper
data set of 62 months of observations, reported that the pulsations
extend above $\unit[50]{GeV}$ with a weak total signal of five photons at a $4\sigma$  significance level.\\

As the brightest persistent source of high-energy $\gamma$-rays with a potential signal in the tens of GeV range,
the Vela pulsar was one of the prime targets in the
commissioning period of the 2012 upgrade of the \hess{} array of imaging atmospheric Cherenkov
telescopes (IACTs), located in the Khomas Highland of Namibia
  ($23^{\circ}16\arcmin18\arcsec$\,S,
  $16^{\circ} 30\arcmin 00\arcsec$\,E, 1800\,m). 
This upgrade, referred to as \hessII{}, consisted of the addition of a $\unit[28]{m}$ equivalent diameter
telescope (CT5) to its core array of four $\unit[12]{m}$ equivalent
diameter telescopes (CT1-4) and was designed to push the energy threshold of the system to below 50 GeV (from
above $\unit[100]{GeV}$), thus bridging the gap with satellite-based
\gr{} instruments. 

Previous observations of the Vela pulsar with the \hess{} array, above a threshold energy of 170~GeV, had only resulted in upper limits  \citep{HessPSRULs2007}.
In this paper we report the
detection of pulsed \grs{} from the Vela pulsar using CT5-only data with the
aim of reaching the lowest accessible energies.
In order to evaluate the telescope performance, for the first time, near its trigger threshold, \fermi-LAT data from the Vela pulsar is analysed and used as reference.
The light curve and its energy dependence are investigated in the 1-80~GeV and
the spectra of its different features are derived using both instruments. Results are subsequently compared and 
their implications on pulsar emission models are discussed. 

The paper is organized as follows: \hessII{} observations and data analysis are presented in 
section \ref{sec:HessIIObservationsAndAnalysis}.  Section \ref{sec:FermiAnalysis} describes the \fermi-LAT data set and analysis.
Light curves and spectra obtained from the two instruments are presented in sections  \ref{sec:LightCurves} and  \ref{sec:Spectra}. They are 
subsequently discussed and compared in section \ref{sec:Discussion}.  The three appendices give details of the 
commissioning of the  new telescope, investigation of systematic errors on spectral fits, and complementary tables. 

\begin{figure*}[t]
\centering \includegraphics[width=8cm]{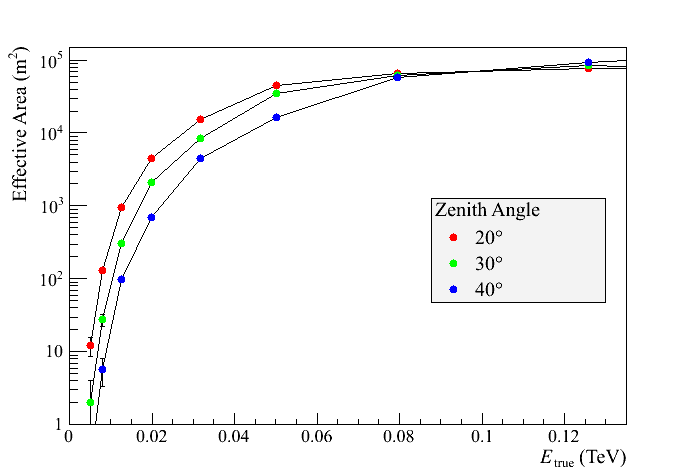}
\centering \includegraphics[width=8cm]{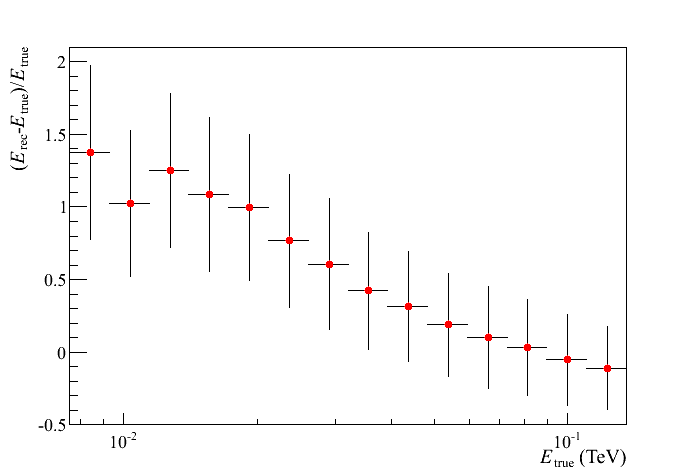}
\caption{{\it Left:} Effective area of \hessII{}-CT5 in monoscopic mode as a function of energy for different zenith angles: the specific analysis used here has 
been developed to yield a large effective area near threshold, i.e. 
$\unit[2.6\times10^2]{m^2}$ at $\unit[10]{GeV}$ and  $20^\circ$ zenith angle. There is a strong dependency of the effective area near the threshold on
zenith angle, e.g. a drop of a factor $\sim$ 10 below $\unit[30]{GeV}$ when comparing $20^\circ$ and $40^\circ$ zenith angles.  
{\it Right:} Distribution of $(E_{\rm rec}-E_{\rm true})/E_{\rm true}$ as a function of $E_{\rm true}$  at $20^\circ$ 
  zenith for a power-law distribution between 5 and 120 GeV with
  index $\Gamma=4$.\ The parameters  $E_{\rm rec}$ and  $E_{\rm true}$ stand for reconstructed and simulated energies, respectively. Error bars show 
  the spread (RMS) of events around the average value. We note that bins are correlated.   
\label{fig:MonoRecoAreaEnergydispersion}}
\end{figure*}

\section{\hessII{} observations and data analysis} 
\label{sec:HessIIObservationsAndAnalysis}
\subsection{Observations}
\label{sec:HessIIObservations}
Vela pulsar observations with \hessII{} were carried out from March to May
2013,  December to April 2014, and February to April 2015.      
A total of 40.3 hours of data were selected based on weather conditions and
instrumental effects  while requiring a zenith angle smaller than
$40^\circ$to  reach a lower energy threshold and better
sensitivity.  The data were split into two sets corresponding to the
commissioning period in 2013-14 (data set I in the following), and normal instrument operation in 2015 (data set II), consisting of 24 and 16.3 hours, respectively. 
Observations were made in wobble mode (\citealt{hess:Crab}) with a source-to-centre of the field-of-view distance of $0.7^\circ$.

\subsection{Data analysis overview} 
\label{sec:HessIIAnalysis}
To test and commission CT5 near its threshold energy,  data from smaller telescopes were discarded 
during the analysis step and a new event reconstruction technique was developed.
Following the raw data reduction using the calibration runs (see \citealt{hess:Crab} for an overall view), 
shower image intensities were obtained after application of a
tail-cuts algorithm with two thresholds, i.e. five and seven photo-electrons (or
p.e.), in order to remove noisy pixels and night sky background
effects.  
The reconstruction algorithm was applied to each image to estimate the shower direction, impact distance, and energy. 
The background rejection is based on image shape parameters and a spatial cut on the reconstructed angle of events with respect to the source position.
Further separation (of signal and background) was obtained in the time domain through selection in phase according to on- and off-phase intervals, which 
are defined \textit{a priori} using the \fermi-LAT light curve.
The overall monoscopic analysis pipeline and the response model of CT5 were validated through a detailed comparison of parameter distributions 
for Monte Carlo-generated (MC) events with those obtained from the observations. The expected distributions were derived using  the \fermi-LAT spectrum as input to simulations. 
Details of this procedure are given in Appendix \ref{sec:CT5Validation}.  

\subsection{Event reconstruction} 
\label{sec:HessIIEventReconstruction}

The event reconstruction performs a non-parametric estimation of the shower properties, i.e. direction,
impact distance, and energy, using a multi-layer perceptron artificial neural network.
The training data set was produced through extensive MC simulations of the overall detection chain,
starting from the generation of electromagnetic showers in the
atmosphere down to the light collection and charge measurement in the
detector \citep{SimTelArray}.  
The image intensity, $Q$, its second order moments \citep{HillasParameters}, i.e. length $l$, width
$w$, together with the angular distance, $d$, of the image
barycentre to the source position, were used as input parameters to the
neural networks.

The reconstruction algorithm for the event direction assumes that the shower direction
projected onto the camera plane lies on the major axis of the image and
its angular distance $d$ from the barycentre is a function of $l$, $w$, and $Q$. 
For sources of known position, for example  pulsars, it is further
assumed that the source lies towards the centre of the field of
view. This provides a better angular resolution at the lowest
energies, at the expense of a higher background. 
To train the estimator for the impact distance, $\rho$, the actual value of
$d$, calculated using the true source position in the field of
view, is used as input in addition to the
above-mentioned image parameters.
The energy estimator relies on the following five parameters: $Q$, $l$, $w$, $d$, and $\rho$. During the training
phase, the true value, $\rho_{\rm true}$, is used to avoid the smearing of the estimator response by the impact distance error.\\

\subsection{Background rejection} 
\label{sec:HessIIBkgRejection}
In addition to the selection through a spatial (angular) cut at the 68\% containment radius, $\rm R_{68}$, 
the background rejection (of non-$\gamma$-ray showers) relies on image shape and 
on estimates of shower physical parameters. 
A multi-variate boosted decision tree (BDT) classifier, based on the TMVA package \citep{TMVA}, is used in the same spirit
as that for the \hess~I array \citep{Becherini2011}. 
During the training step, MC simulations of $\gamma$-ray induced images and
real off-source data are used as signal and background inputs,
respectively.  In addition to $l$ and $w$, physical
parameters of the shower are obtained thanks to a 3D Gaussian-model fit of the
corresponding photo-sphere \citep{Marianne3D2006}, and are used to improve the
discrimination power. These consist of the shower length, width, and the depth of its maximum in the atmosphere.  
During the 3D fit of monoscopic events, the shower direction and impact distance are fixed to values obtained 
by the event reconstruction algorithm.

For the BDT-response parameter, $\zeta$, uni-modal distributions are
obtained both for signal and background training samples, and test
samples are checked for their compatibility with the training samples, to exclude over-training artifacts.
Both the training and test samples of the signal consist of MC-generated \grs{}. 

\subsection{Performance}
\label{sec:HessIIPerformance}
In order to reach a low-energy detection threshold, and given the expected very soft energy spectrum of the Vela pulsar (see sect. \ref{sec:FermiAnalysis}),
analysis cuts were optimized such as to yield a large effective area
in the 10-20 GeV range, at the cost of a reduced $\gamma$-background separation. 

The main analysis cut configuration (Cuts~I) is based on an image intensity cut, $Q> Q_{\rm min}=30$ p.e., 
and a BDT discrimination cut, $\zeta > -0.1$. The selected events exhibit a 68\% containment radius $\rm R_{68} = \unit[0.3]{^\circ}$ for the reconstructed direction. 
The reconstructed energy, $E_{\rm rec}$, shows a large bias near the detection threshold, decreasing with increasing energy (see Fig.~\ref{fig:MonoRecoAreaEnergydispersion}, right panel).

This set of cuts provides a background rejection $\rm B_{\rm rej}$=96\%  and
an effective area $\rm A_{\rm eff} \sim \unit[4.5\times10^3]{m^2}$ at $\unit[20]{GeV}$ at a zenith angle of $20^\circ$ (see Fig.~\ref{fig:MonoRecoAreaEnergydispersion}, left panel).
A second configuration with a higher energy threshold, (Cuts II), consists of a two times larger image intensity cut, $Q_{\rm min}=60$ p.e., together with 
a tighter spatial cut, $\rm R_{68} = \unit[0.224]{^\circ}$, thanks to the improvement of the point spread function.  
The resulting rejection is $\rm B_{\rm rej}$=98.4\%, for $\rm A_{\rm eff} \sim \unit[1.3\times10^3]{m^2}$ at $\unit[20]{GeV}$.

\subsection{Timing and phase selection}
\label{sec:HessIITiming}
Event time stamps are provided by the central trigger system of the \hess{} array with a long-term 
stability of better than 2 $\mu$s. This stability has been obtained thanks to a GPS receiver and various software corrections 
of the timing in the array (e.g. leap seconds and polar motion) \citep{TriggerFunk2004}.      
The pulsar phase corresponding to each event is calculated using the \texttt{Tempo2} package \citep{TempoGeneral}.  Arrival times are 
transformed to the solar system barycentre and the phase of each event is computed using an ephemeris based on radio data from the Parkes Radio Telescope. 
The ephemeris is valid in the range \texttt{MJD}~54175.52-57624.20 (with fiducial phase reference, \texttt{TZRMJD} = 55896.55) 
with a precision of better than 1 milli-period ($91 \mu $s), degrading 
to $\sim 10$~milli-periods around the glitch at \texttt{MJD}~56555.8.  We note, however, that  there were no \hessII{} data taken in the vicinity of the  glitch.

The search for pulsed signals is performed using a maximum likelihood-ratio test \citep{LiAndMa83} 
on counts extracted from \textit{a priori} defined  on- and off-phase intervals (see section \ref{sec:FermiLightCurve}) and by applying the H-test periodicity  
test \citep{deJager89}. The latter makes no prior assumptions about the light-curve model.

\begin{figure*}[t]
\centering
\includegraphics[width=8.5cm]{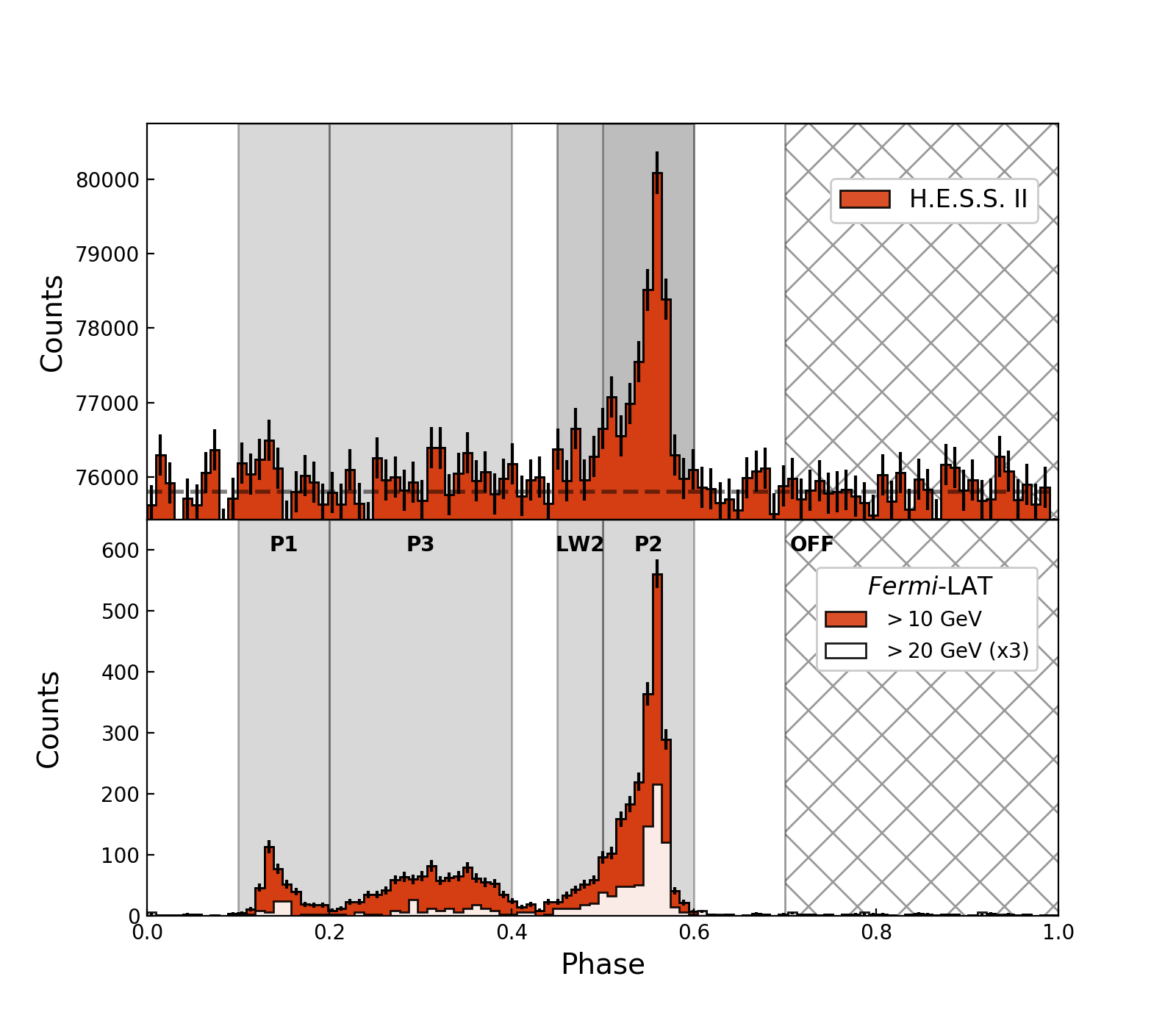}
\vspace{0.3cm}
\includegraphics[width=8.cm]{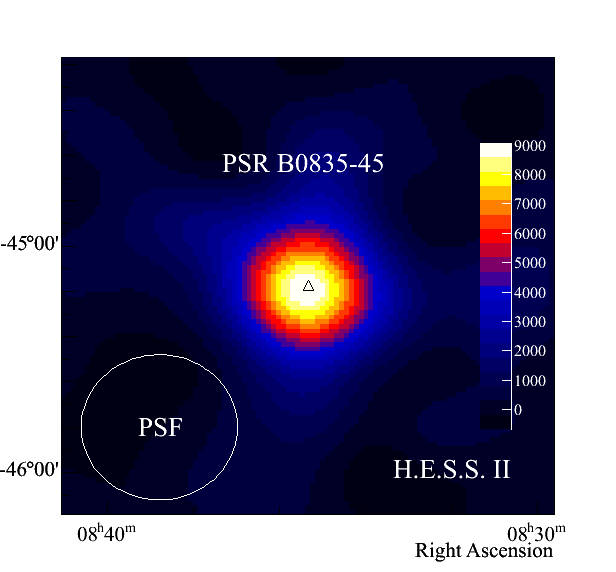}
\caption{{\it Left:}  \gr{} phasogram of the Vela pulsar obtained using 40.3 h of \hessII-CT5 data with the Cuts~I analysis configuration (top panel) 
and 96 months of \fermi-LAT data above 10 and 20~GeV (bottom panel). The dashed line on the CT5 phasogram shows the level of the background estimated in the [0.7-1.0] phase range.
{\it Right:} Gaussian-smoothed excess map ($\sigma=0.15^{\circ}$) for the CT5 data in the P2 phase range, where the on and off maps are made  
after selection of events in on- and off-phase intervals defined as [0.5-0.6] and [0.7-1.0], respectively. The black triangle indicates the position of the pulsar.
\label{fig:CombinedHESSFermiLC}}   
\end{figure*}

\subsection{Spectral derivation}
\label{sec:HessIISpectrumMethod}
The energy spectra are derived using a  maximum
likelihood fit within a forward-folding scheme, assuming \textit{a priori} spectral models~\citep{Piron2001}.
Two sets of 
instrument response functions (IRFs) were used for each cut configuration to account for 
different optical efficiencies of CT5 in data sets I and II.
These IRFs were computed through extensive MC simulations as a
function of the energy, zenith, and azimuthal angles of the telescope
pointing direction, the impact parameter of showers, and the
configuration of the telescope for each observing period.  
The pipeline was tested and validated by simulating   
150 spectra of $\gamma$-rays with a steep power law (with index $\Gamma = 4.0$) 
added to background events such as to reach a signal-to-noise ratio similar to that of real
data. 
In this way it is possible to estimate the energy threshold, $E^{\rm safe}_{\rm rec}=20$~GeV in reconstructed units, 
defined as the energy above which the bias in the reconstructed parameters, due to the uncertainties in the IRFs, remains negligible
as compared to the statistical errors  
(see Appendix \ref{sec:HESSIISystematicErrors} for more details on this point).

\section{\fermi-LAT data analysis}
\label{sec:FermiAnalysis}
\fermi-LAT data were used to derive the expected signal above $\unit[10]{GeV}$ and to define the phase ranges 
of interest subsequently.
The analysis was performed using 96 months of data acquired from August 4, 2008, to July 26, 2016. 
Events were selected in an energy range of $\unit[100]{MeV}$ to $\unit[500]{GeV}$ in the P8 Source class (event class = 128, event type =3) within a 
region of interest (ROI) of $\unit[10]{^{\circ}}$ radius around the position of the Vela pulsar;
\texttt{P8R2\_SOURCE\_V6} IRFs 
were used.
In order to reduce contamination
by $\gamma$-rays from Earth's limb, only $\gamma$-ray events 
with reconstructed zenith angles smaller than $\unit[90]{^{\circ}}$ were selected.
For the specific study of the energy range overlapping with that of \hess~II, events were selected above $\unit[10]{GeV}$ and the ROI was limited to 
$\unit[5]{^{\circ}}$.  
To compute the pulsar phase, selected events were folded 
using the \texttt{Tempo2} \fermi{} plug-in \citep{FermiPluginAndGTP} and the same ephemeris as that used for the \hessII{} data (see section \ref{sec:HessIITiming}).
To generate the light curves, an additional selection cut was applied on the angular distance of each photon to the pulsar 
position, where $\theta_{\rm max}=0.8^\circ$. This value corresponds to the 68\% and 95\% containment radii of the \fermi-LAT at 1 and 10~GeV, respectively, and allows 
us to retain a large number of highest energy photons, while limiting the background in the $1-10$~GeV range.  

Spectral analysis was performed with \texttt{gtlike}, with 
Galactic diffuse emission model, \texttt{gll\_iem\_v06.fits}, and
isotropic diffuse model, \texttt{iso\_P8R2\_SOURCE\_V6\_v06.txt}.
All sources from the \fermi-LAT third source catalogue (3FGL) \citep{2015ApJS..218...23A}, within a region of
$\unit[20]{^{\circ}}$ radius centred on the pulsar position were added to
the source model, while parameters for sources outside the ROI were
fixed during the fit.   
Finally, energy dispersion correction was enabled during the analysis to minimize systematic effects. 
More details are given in Appendix \ref{sec:FermiSystematicErrors}.

\section{Light curves}
\label{sec:LightCurves}
\subsection{\fermi-LAT}
\label{sec:FermiLightCurve}
The \fermi-LAT phasogram above $\unit[10]{GeV}$ of the Vela pulsar is shown in the bottom left panel of Fig.~\ref{fig:CombinedHESSFermiLC}.
It contains a total of 3817 events including a background of 190 events estimated using the phase interval [0.7-1.0].
This range, initially defined as the off-pulse interval in \cite{fermifirstvela}, was restricted to [0.8,1.0] in \citet{Abdo2010Vela} to limit 
contamination of background by the low-energy (i.e. < 1~GeV) trailing edge of P2.  Given the focus of this study above the CT5 energy 
threshold, i.e. well above 1~GeV, the initial background phase range was retained.      
On-pulse phase ranges were defined for the three main features of the pulsar light curve as follows: [0.1-0.2] for the first
peak, P1, [0.2-0.4] for the bridge, P3, [0.45-0.5] for the leading wing of the second peak, labelled LW2,  and [0.5-0.6] for P2 itself.
The latter is the most prominent peak in the phasogram, exhibiting 1977 excess events after subtraction of 19 background counts. Peaks   
P1, LW2 and P3 show lower intensities with excess counts of 382, 227, and 953, respectively. 
All three peaks are still present for energies higher than 20~GeV;
P1, P2, and P3 have excess counts/significance 
levels of 21/5.1$\sigma$, 228/23.2$\sigma$ and 43/6.3$\sigma$, respectively. Peak LW2 is also present 
above 20~GeV with 29 excess counts at 8.3$\sigma$. We note the high significance of LW2 above 20~GeV because its intensity relative to P2, LW2/P2$\sim 15$\% does not decrease 
with increasing energy. This is in contrast to the P1/P2 ratio, which 
drops very quickly, or even to the P3/P1 ratio, which exhibits a smaller decrease. This hints at a harder spectrum for LW2, as compared to the two other features. 
This point is investigated further in sections \ref{sec:HessIILightcurve}, \ref{sec:FermiSpectra}, and \ref{sec:Discussion}.        

\begin{table}[htb]
\caption{Number of events, excess counts (Excess),  and significance (Sig) derived from the \hessII{} light curve for 
the four features of the Vela pulsar as defined by their phase intervals. 
Results are given for the two analysis configurations, Cuts I and II. The last row gives  
the number of events in the Off region.\label{tab:HessVelaStatistics}}             
\centering              

\begin{scriptsize}  
\begin{tabular}{l c | c c c |c c c}        
\multicolumn{2}{c|}{Configuration}   & \multicolumn{3}{c|}{Cuts I}     & \multicolumn{3}{c}{Cuts II}\\
\hline\hline                        
 & Interval  & Events &  Excess & Sig & Events &  Excess & Sig \\    
\hline                        
P1 &  [0.1-0.2] & 767253   & 1574  & 1.6   & 324603 & 967  & 1.5 \\      
LW2&  [0.45-0.5]& 385270   & 2431  & 3.6   & 163767 & 1949 & 4.5 \\      
P2 &  [0.5-0.6] & 781514   & 15835 & 15.6  & 330626 & 6990 & 10.6 \\      
P3 &  [0.2-0.4] & 1534381  &  3023 & 1.9   & 649759 & 2487 &  2.4 \\      
\hline                                   
OFF&  [0.7-1.0] & 2297037  & -     & -     & 970908 & -     &  - \\
\hline                                   
\end{tabular}
\end{scriptsize}
\end{table}

The above-mentioned phase ranges for P1, LW2, P2, P3, and for background estimation are 
used when searching for signals with \hessII{} in the following section. 

\begin{table*}[htb]
\caption{Best-fit parameters of an asymmetric Lorentzian model for the P2 peak, obtained from the \fermi-LAT and \hessII{} data in different energy ranges, 
as part of a three-component function fit to the Vela pulsar light curve 
(see text). Parameters are the fitted position, $\phi_{\rm P2}$, 
the leading (or inner) edge width, $\sigma_{\rm L}$, and the trailing (or outer) edge width,  $\sigma_{\rm T}$.
An additional Gaussian component with position, $\phi_{\rm G,\,P2}$, and width,  $\sigma_{\rm G,\,P2}$,  is included for the fit to the \hessII{} full energy range light curve.
The energy ranges and corresponding average energies are also given (see text). \label{tab:P2Charateristics}}            
\begin{center}
\centering                          
\begin{tabular}{c c c c c c }        
\hline\hline                 
Instrument & Energy range &  $\langle E_{\rm true} \rangle$  & $\phi_{\rm P2}$  &  $\sigma_{\rm L}$  &  $\sigma_{\rm T}$  \\    
           & (GeV)        & (GeV)                    &  (phase units)  & (phase units)& (phase units) \\    
\hline                        
   \fermi-LAT & 1-3  & 1.7  & 0.5648 $\pm$ 0.0001 & 0.0327 $\pm$ 0.0002 & 0.0080 $\pm$ 0.00008\\      
   \fermi-LAT & 3-10 & 4.8  & 0.5653 $\pm$ 0.0002 & 0.0323 $\pm$ 0.0004 & 0.0056 $\pm$ 0.0001\\      
   \fermi-LAT & 10-20 & 13  & 0.5650 $\pm$ 0.0005 & 0.025 $\pm$ 0.001 & 0.0038 $\pm$ 0.0003\\      
   \fermi-LAT &  >20  & 28  & 0.565  $\pm$ 0.001  & 0.017  $\pm$ 0.002  & 0.0029 $\pm$ 0.0008\\      
\hline                                   
\vspace{0.12cm}
\hessII{}  & $\sim 10$-33 & 19 & $0.564^{+0.001}_{-0.001}$       & $0.019^{+0.003}_{-0.002}$ & $0.006^{+0.001}_{-0.001}$\\      
\vspace{0.12cm}
\hessII{}  & $\sim 20$-80 & 42& $0.5697^{+0.0005}_{-0.0011}$ & $0.031^{+0.006}_{-0.005}$ & $0.0007^{+0.0015}_{-0.0007} $\\      
\vspace{0.1cm}
\hessII{}  & $\sim 10$-80& 31& $0.5684^{+0.0007}_{-0.0013}$    & $0.027^{+0.003}_{-0.003}$ & $0.002^{+0.0014}_{-0.0008}$\\      
\hline                                   
\hline                                   
\hessII{} & $\sim 10$-80 & 31& ${0.5691^{+0.0006}_{-0.0009}}$   &  ${0.030^{+0.004}_{-0.003}}$ &  ${0.002^{+0.001}_{-0.0009}}$\\      
\cline{4-5}
\multicolumn{1}{l}{} & \multicolumn{2}{c}{}  & $\phi_{\rm G,\,P2}$  &  ${\sigma_{\rm G,\,P2}}$  & \\
\cline{4-5}
\multicolumn{1}{l}{} & \multicolumn{2}{c}{} & ${0.5569^{+0.0006}_{-0.0007}}$    & ${0.0022^{+0.0008}_{-0.0007}}$ &\\
\hline                                   
\end{tabular}
\end{center}
\end{table*}

\subsection{\hessII{}}
\label{sec:HessIILightcurve}
The phasogram of the Vela pulsar for the whole data set, 
obtained in monoscopic mode with the main analysis configuration, Cuts~I, is shown in the top left panel of 
Fig.~\ref{fig:CombinedHESSFermiLC}. 
Using the H test for periodicity on the full phasogram range yields a significance of $>17.5\sigma$. 
A simple maximum likelihood-ratio test \citep{LiAndMa83} within the predefined phase range for P2, [0.5-0.6],
results in an excess of 15835 events, at a significance level of $15.6\sigma$. The background, evaluated in the [0.7,1.0] phase interval,  
amounts to 765679 events after normalization (see Table~\ref{tab:HessVelaStatistics}). 
The excess map is shown in the right panel of Fig.~\ref{fig:CombinedHESSFermiLC}. 
Data set~I was used for validation of the analysis pipeline and of the CT5 response model (see Appendix~\ref{sec:CT5Validation}) and as such
was subject to few trials ($<$10). Its analysis with the Cuts~I configuration  yields an excess of 9789 events at a significance level of $12.4\sigma$.  
This high level of significance makes any correction for the trials factor irrelevant.
Analysis with the same configuration of data set II yields an excess of 6047 events at $9.4\sigma$.  
While the phase intervals P1 and P3 show positive excess counts, they are not significant based on a simple likelihood-ratio 
test (whether analysed individually or jointly); see Table~\ref{tab:HessVelaStatistics}.  
The figures do not improve with a phasogram model-based likelihood-ratio test 
(see section~\ref{sec:P2properties}) nor  with the higher threshold analysis configuration, Cuts~II.
This is in contrast with the leading wing of P2, i.e. LW2. Indeed, while LW2 shows an excess of 2431 events at  $3.6\sigma$ with the low-threshold analysis configuration (Cuts~I),
its significance reaches $4.5\sigma$ with Cuts~II, with a corresponding excess of 1949 events. This reinforces the hint of a hard spectrum for LW2 found in the
\fermi-LAT data in section~\ref{sec:FermiLightCurve}.

The true-energy distribution of events in excess in the P2 phase range was 
derived using MC simulations (see details in Appendix~\ref{sec:CT5Validation}) and is shown in Fig.~\ref{Fig:Theta2AndEnergyDistribution}, right panel.
The average true energy of the excess is 31~GeV, 40\% of events lie below 20~GeV (i.e. $\sim$6400 events out of  the total excess), 
36\% ($\sim$5400) are in the 10-20~GeV range, and still 6\% ($\sim$1000) below 10~GeV.

\begin{figure}[thb]
\centering
\includegraphics[width=8.4cm]{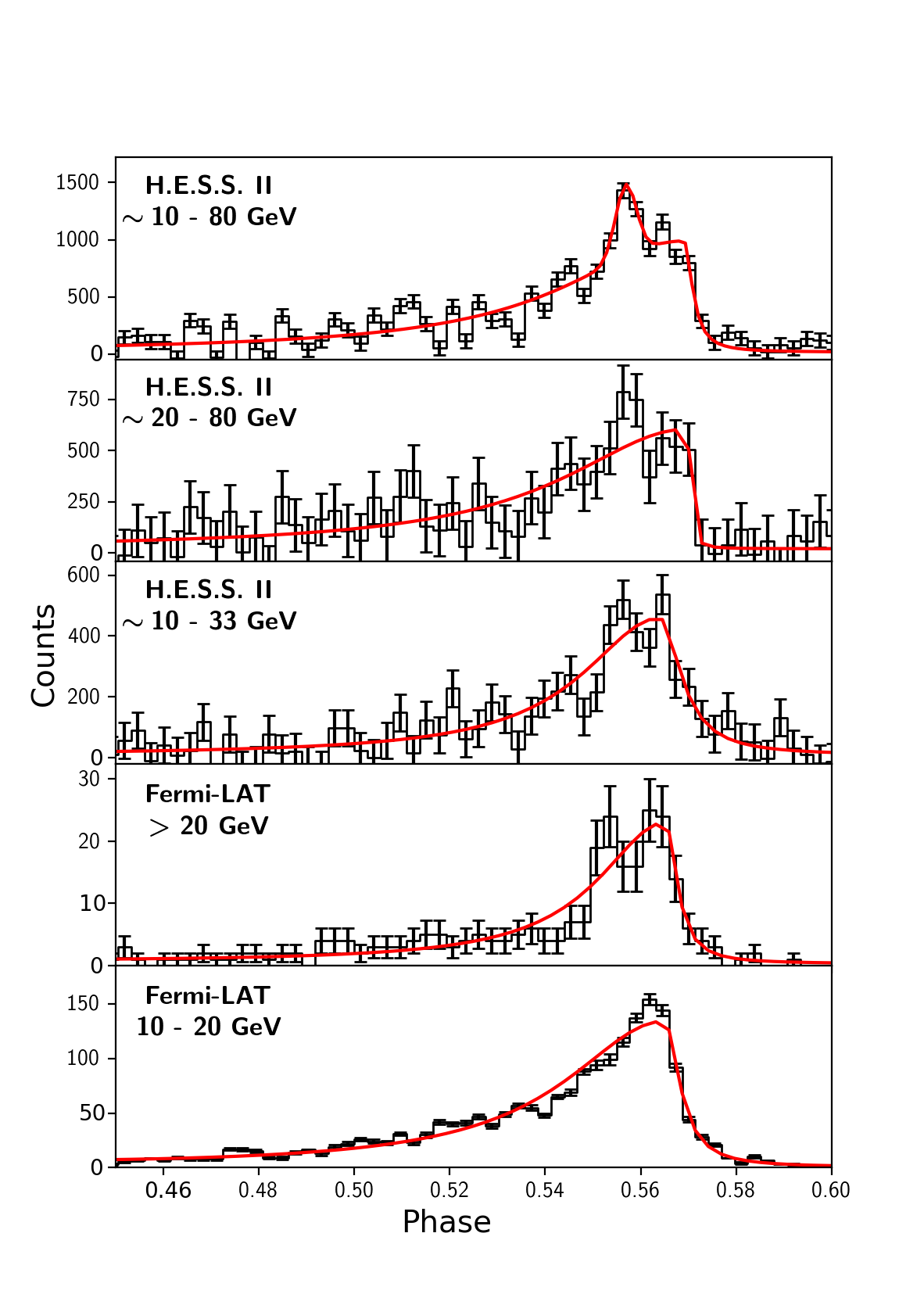}
\caption{ \gr{} phasogram of the Vela pulsar around P2 in the phase range [0.45, 0.6], in different energy bands from 10 to 80~GeV, as obtained from the \fermi-LAT and 
\hessII{} data. The red curve represents the asymmetric Lorentzian form fitted to data, except for the top-most panel where an additional Gaussian component 
is included in the fit to the \hessII{} full energy range light curve (see Table~\ref{tab:P2Charateristics}). 
The background has been subtracted for the \hessII{} light curves. The bin width for all histograms is 0.25 ms.
\label{fig:HESSFermiLC2EnergyBands}}   
\end{figure}

\subsection{Light curve evolution with energy as seen with both instruments}
\label{sec:P2properties}

In this section the light curves obtained with the \fermi-LAT are compared to those derived from \hessII{} data, based on the same ephemeris (see sections 
\ref{sec:HessIITiming} and  \ref{sec:FermiAnalysis}).   
As the \hessII{} data span only a subset of the \fermi{} observation period (i.e. 96 months), the stability of the LAT phasograms corresponding to the  
overlapping period and the overall data set was checked. This showed no measurable variation and hence systematic effects are expected to be negligible 
when comparing the two instruments light curves.

To characterize the \fermi-LAT phasograms, following \citet{Abdo2010Vela}, two asymmetric Lorentzian functions and a log-normal function were used 
in a joint unbinned maximum likelihood fit to P1, P2, and P3, respectively. 
The same functions were used to fit the \hessII{} data. However, given the low significance of P1 and P3, all of their parameters 
except the amplitude were fixed to values obtained above 10~GeV with the \fermi-LAT.
Including in this manner both P1 and P3 (only P1 or P3) in the fit  improves the likelihood 
at a level of 3.4$\sigma$ (2.8$\sigma$ and 1.7$\sigma$, respectively), when compared to a P2-only fit.     

In order to study the evolution of the light curve as a function of energy, 
data of both instruments were subdivided into energy bands as follows: 1-3, 3-10, 10-20,  and  >~20~GeV for the LAT; and 
$\sim$10-33~GeV and $\sim$20-80~GeV for CT5.
The former CT5 band is obtained by selecting events with $E_{\rm rec}<40$GeV and the latter by using the higher threshold Cuts~II analysis configuration.
The approximate lower bounds and overlap between these two energy bands are caused by the migration of events from lower to higher energies owing to the                        
large bias and dispersion in energy reconstruction (see section \ref{sec:HessIIEventReconstruction}, and Fig.~\ref{fig:MonoRecoAreaEnergydispersion}, right panel). 
The upper bound of 80~GeV corresponds to the average true energy in the last significant bin of the spectrum (discussed in section \ref{sec:HessIISpectra}).  

\begin{table*}[hbt]
\caption{Phase-averaged (PA) and phased-resolved spectra parameters for the Vela pulsar obtained with 96 months of \fermi-LAT data using the 
exponentially cut-off power-law hypothesis (ECPL) above 100~MeV, and a simple power law (PWL) for energies $>10$~GeV.
 The flux normalization, $\Phi_0 $, is given in units of $ 10^{-9}$MeV$^{-1}$cm$^{-2}$ s$^{-1}$ at the reference energy $E_0=1$~GeV for the ECPL fit, and 
in units of $10^{-9}$TeV$^{-1}$cm$^{-2}$ at  $E_0=25$~GeV for the PWL model.   
Phase ranges of the various features are defined in Table~\ref{tab:HessVelaStatistics}\label{tab:FermiSpectra}.}             
\begin{center}
\centering                          
\begin{minipage}{\textwidth}
\begin{center}
\begin{footnotesize}  
\begin{tabular}{l| c c c c | c c }        
\hline\hline                 
\multicolumn{1}{l|}{Model}   & \multicolumn{4}{c|}{ECPL ($>100 $MeV) : $\Phi_0 \left({E}/{E_0}\right)^{-\Gamma} \exp\left[-\left({E}/{E_\mathrm{c}}\right)^b\right]$) }  & \multicolumn{2}{c}{PWL ($>10$ GeV) : $\Phi_0 \left({E}/{E_0}\right)^{-\Gamma}$}\\
\hline
 & $\Phi_0$  & $\Gamma$ & $b$ &  $E_c$ (MeV) & $\Phi_0$ & $\Gamma$ \\    
\hline                        
PA &  17.64 $\pm$ 0.02 &  0.913 $\pm$ 0.003 & 0.439 $\pm$ 0.001 & 143 $\pm$ 2 & --  & -- \\      
P1 &  4.36$\pm$ 0.08   &  1.086 $\pm$ 0.005 & 0.468 $\pm$ 0.002 & 164 $\pm$ 4 &  3.48 $\pm$0.58  &  5.24 $\pm$ 0.27  \\      
P2 &  8.28$\pm$ 0.13   &  0.890 $\pm$ 0.004 & 0.385 $\pm$ 0.001 & 78 $\pm$ 2  &  40.3 $\pm$ 1.8  &  4.10 $\pm$ 0.08\\      
LW2&  4.97$\pm$ 0.22   &  0.916 $\pm$ 0.013 & 0.523 $\pm$ 0.007 & 385 $\pm$ 24&  4.84 $\pm$ 0.73 &  4.17 $\pm$ 0.27 \\
P3 &  3.70$\pm$ 0.10   &  0.71 $\pm$ 0.07   & 0.51 $\pm$ 0.03 & 254 $\pm$ 89  &  12.7$\pm$1.2  &  5.05 $\pm$ 0.16\\      
\hline                                   
\end{tabular}
\end{footnotesize}
\end{center}
\end{minipage}
\end{center}
\end{table*}

Fit results for P2  (given in Table~\ref{tab:P2Charateristics}) confirm those obtained by \citet{Abdo2010Vela} in the  1-20~GeV range, i.e.
the fitted position of P2, $\phi_{\rm P2}=0.565$, shows no variation, while its width decreases with increasing energy mainly owing to the sharpening of 
the outer (trailing) edge,  $\sigma_{\rm T}$, up to 10~GeV, and then also because of a decrease of its inner (leading) edge width, $\sigma_{\rm L}$.  
We note, however, that the fit residuals are large, implying that the asymmetric Lorentzian model 
is not sufficient to describe the peak (see e.g. the 10-20~GeV fit in Fig.~\ref{fig:HESSFermiLC2EnergyBands}).
Alternatively, the use of a Gaussian kernel density estimator (KDE) results in similar 
conclusions, except that the estimated peak position, $\phi_{\rm KDE,\,P2}^{\rm LAT}=0.561$, is offset towards earlier phases, as it  
corresponds to the maximum of the peak distribution.
Above 20~GeV in the \fermi-LAT data and in the first \hessII{} energy band,  
the fitted position for P2 remains unchanged and the narrowing of both of its edges continues.
While the KDE estimate of the position for the LAT is also stable, that of \hessII{}, $\phi_{\rm KDE,\,P2}^{\rm HESS }=0.557$, is at variance with the former and with 
the fitted position, $\phi_{\rm P2}^{\rm HESS }=0.564$. This variation might be attributed to the onset of a second component in the phase range [0.550, 0.558], which is also
apparent in the second and highest \hessII{} energy band (see Fig.~\ref{fig:HESSFermiLC2EnergyBands}).  
To test this hypothesis, a Gaussian component was added to the asymmetric Lorentzian function and fitted to data. While the limited statistics yield only a marginal 
evidence for such a component in each of the energy bands fitted separately (i.e. $\lesssim 2\sigma$), a significance level of $3.4\sigma$ is obtained 
for the full energy range (shown on the top-most panel of Fig.~\ref{fig:HESSFermiLC2EnergyBands}). The additional component exhibits a Gaussian width of $\sigma_{\rm G,\,P2}=0.002$ and its 
fitted position, $\phi_{\rm G,\,P2}=0.557$, coincides with the KDE result, as expected. 

{In the full range and in the highest energy band, the fitted position of P2 moves to later phases by $\sim$5 milli-periods, $\phi_{\rm P2}\simeq 0.569$, 
its trailing edge continues to sharpen, narrowing down to a width compatible with zero,  whilst its leading edge width, $\sigma_{\rm L}$,  widens.}
As the peak maximum stays stable, the change in the fitted position is mainly caused by the strong sharpening of the P2 trailing edge rather than by a shift of the peak as a whole.    
Combined with the hardening of LW2 in the tens of GeV range (see sections \ref{sec:FermiLightCurve}, 
\ref{sec:FermiSpectra}, and \ref{sec:HessIISpectra} below), this is possibly the cause of the larger value fitted for $\sigma_{\rm L}$.

The fitted parameters for P1 and P3 (given in Table~\ref{tab:FermiP1P3FitResults}) and their evolution with increasing energy, up to 20~GeV in the LAT data,
are also in line with the results reported in \citet{Abdo2010Vela}, namely: (i) no measurable change in the P1 position; (ii) a sharpening of 
its leading (or outer) edge and an initial increase of its trailing edge width above 3~GeV before a decrease; and
(iii) a pronounced movement towards later phases of the P3 centroid accompanied by 
a narrowing of its width. We note important fit residuals here as was the case for P2, in particular close to the
maximum of P1. The KDE estimated position of P1 shows, as in the case of P2, an offset $\Delta_{\phi} (\rm P1)=\phi_{\rm P1}^{\rm KDE}-\phi_{\rm P1}$ with respect to
the fitted one, except that $\Delta_{\phi} (\rm P1)$ is positive here and increases with increasing energy, varying   
from +3 to +6 milli-periods in the 1 to 20 GeV range.
Above the latter energy, both $\phi_{\rm P1}=0.158\pm0.003$ and $\phi_{\rm P1}^{\rm KDE}=0.148 \pm 0.004$ move
towards higher phase values, but the scarcity of statistics forbids any firm conclusion. The same limitation holds for the widening of P3, i.e. the 
increase of $\sigma_{\rm P3}$=0.157$\pm$0.007 in the 10-20~GeV band to 0.39$\pm$0.10 above 20~GeV.             

\section{Spectra}
\label{sec:Spectra}

\subsection{\fermi-LAT}
\label{sec:FermiSpectra}
Phase-averaged (PA) and phase-resolved spectra were fitted first above 100~MeV,
assuming a power law with an exponential cut-off (ECPL,  ${\rm d}N(E)/{\rm d}E = N_0 \left({E}/{E_0}\right)^{-\gamma} \exp\left[-\left({E}/{E_\mathrm{c}}\right)^b\right]$).
Results, summarized in Table \ref{tab:FermiSpectra}, show that the best-fit values of $b$ differ significantly from unity. There is hence a clear preference
for a sub-exponential cut-off for the PA spectrum, as already shown by \citet{Abdo2010Vela}, but also for the phase-resolved
spectra of P1, P2, LW2, and P3. These more precise determinations are obtained thanks to the large event statistics accumulated 
with the additional exposure, and to the improved performance of the \texttt{P8} data processing software.    
In a second step, the spectra of the four light curve features were derived above $\unit[10]{GeV}$, i.e. in the energy range overlapping with that of \hess~II. 
A simple power law was assumed that was independent from the lower energy part of the emission.
For the strongest peak at these energies, P2,  a spectral index\footnote{In fact, the local slope of the spectrum, given the reduced energy range studied here.} of 
$\Gamma^{\rm LAT} = 4.10 \pm 0.08^{\rm stat} \pm0.1^{\rm sys}$ best fits the data with   
a normalization $\Phi_0=40.3\pm1.8^{\rm stat} \pm0.5^{\rm sys} \times 10^{-9}$ {TeV}$^{-1}$cm$^{-2}$ s$^{-1}$, 
at a reference energy $E_0=25$ GeV (see Fig.~\ref{Fig:Spectrum}).
The leading edge of P2, LW2 shows a spectrum as hard as that of P2, while both P1 and 
P3 exhibit steeper spectra with indices of $ 5.24 \pm 0.27$ and $ 5.05 \pm 0.16$, respectively.

\begin{figure}[h]
\centering
\includegraphics[width=8.4cm]{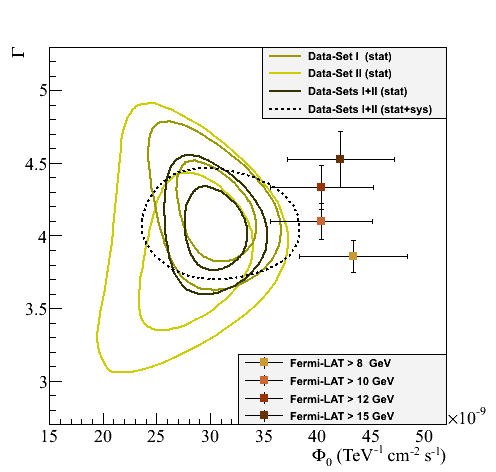}
\caption{
Contours at one and two standard deviations shown for the fitted differential flux at 
25 GeV, $\Phi_0$, and the power-law index, $\Gamma$  for \hessII{} data sets I, II, and their combination.
The dashed curve shows the 1$\sigma$ statistical contour for the overall data set combined quadratically with systematic errors, 
$\delta_{\Phi_0}^{\rm sys}=-$20\%/$+$25\%, $\delta_\Gamma^{\rm sys}=-0.2$/$+0.3$. 
Best-fit values to \fermi-LAT data above  $E_{\rm Thresh}^{\rm LAT}=$~8, 10, 12, and 15~GeV are shown as crosses including 
both statistical and systematic errors, combined quadratically as well. 
The systematic effects on the flux normalizations due to the uncertainty on absolute energy scale of \hessII{} 
and of the \fermi-LAT are not included in the plot. An upper limit of 8\% on the relative shift in the absolute energy scales of the two instruments 
can be inferred based on the deviation of the flux normalization values (see text).
Spectral indices are compatible within errors for all thresholds; the best agreement is 
obtained for $E_{\rm Thresh}^{\rm LAT}=$~10~GeV. 
\label{Fig:Contours}}
\end{figure}

To test for any measurable curvature in the tail at high energies of the P2 spectrum, the simplest quadratic form, i.e. a 
log parabola (LPB, ${\rm d}N(E)/{\rm d}E  = \Phi_0 \left({E}/{E_0}\right)^{-\Gamma-\beta\ \ln({E}/{E_0})}$),
was also fitted to the data above $\unit[10]{GeV}$.  A likelihood-ratio 
test between the power law and the curved model hypotheses favours the latter at a significance level of $S_{\rm LPB}= 3.3\sigma$.
A study of systematic errors due to uncertainties in the model of bright nearby sources (mainly the Galactic plane; see Appendix \ref{sec:FermiSystematicErrors}) 
shows that the best-fit values of parameters for the PL and LPB models ($\Gamma_{\rm LPB} = 4.3 \pm 0.13 $, $\beta = 0.7\pm 0.2$) are stable and 
that  $S_{\rm LPB}$ varies mildly between 3.1$\sigma$ and 3.5$\sigma$.
The impact of the curvature on the power-law index fit to P2 was investigated by selecting data above 
several  energy thresholds, i.e. 8, 12, 15, 20, and 30~GeV.
As could be expected, the index varies,  
ranging from $\Gamma_{8~\rm GeV} = 3.86 \pm 0.05$ above 8~GeV, to $\Gamma_{15~\rm GeV}=4.55 \pm 0.17$ above 15~GeV (see  Fig.~\ref{Fig:Contours}), 
up to $\Gamma_{20~\rm GeV}=4.80 \pm 0.30$ and $\Gamma_{30~\rm GeV}=5.38 \pm 0.78$ for thresholds of 20 and 30~GeV, respectively.   
The log parabola model was also fitted to data for the different thresholds. While the LPB best-fit values do not show 
any significant  change up to 15~GeV, the significance of the curvature, $S_{\rm LPB}$, attains a large value of 7.3$\sigma$ above 8~GeV, 
decreasing to $1.9\sigma$ for 12~GeV and to below $1\sigma$ above 15~GeV.  
This is expected because of the progressive lack of event statistics.   

Compared to P2, LW2 shows an opposite behaviour, i.e. the LPB fit results in a convex curve, where $\Gamma_{\rm} = 4.08 \pm 0.20 $, $\, \beta =-0.60\pm 0.14$, and 
$S_{\rm LPB} =2.4$$\sigma$, 
and a power-law fit above 20~GeV gives $\Gamma_{20~\rm GeV} = 2.80 \pm 0.45$, 
suggesting a hardening as a function of energy. The fit of a broken power-law model (BPL; 
${\rm d}N(E)/{\rm d}E \propto \left({E}/{E_b}\right)^{-\Gamma_1}$ if (E $<E_b$); else $\propto  \left({E}/{E_b}\right)^{-\Gamma_2}$) 
results indeed  in indices $\Gamma_{1\,\rm} = 4.37 \pm 0.24 $ and  
$\Gamma_{2\,\rm} = 1.37 \pm 0.64 $, where the break energy $E_b=50.2 \pm9.5$~GeV.  
The BPL is favoured, however,  only at  $S_{\rm BPL} =2.3$$\sigma$.
This point is further investigated in section \ref{sec:PulsedemissionAbove80GeV}.

\subsection{\hessII{}}
\label{sec:HessIISpectra}

Data were selected with the main analysis configuration, Cuts~I (see section \ref{sec:HessIIEventReconstruction}), 
and according to the definitions for P2 and off-phase intervals given in Table \ref{tab:HessVelaStatistics}.     
The fit of a power law to the overall data set above $E^{\rm safe}_{\rm rec}=20$~GeV results in an index, 
$\Gamma_{\rm HESS}^{\rm CI} = 4.06 \pm 0.16^{\rm stat}$, a normalization,
$\Phi_0^{\rm HESS}=30.6\pm1.9^{\rm stat} \times 10^{-9}$ {TeV}$^{-1}$cm$^{-2}$s$^{-1}$, at the reference energy, $E_0=25$~GeV, and with
decorrelation energy, $E_{\rm d}=21.5$ GeV. 
The systematic uncertainties on normalization $\delta_{\Phi_0}^{\rm sys}=-$20\%/$+$25\%, and on index $\delta_\Gamma^{\rm sys}=-0.2$/$+0.3$,
are discussed in Appendix~\ref{sec:HESSIISystematicErrors}, where it is shown, in particular, that data sets I and II yield compatible results when fitted independently (see Fig. \ref{Fig:Contours}), 
but a large deviation, $\delta_{\Phi_0}^{\rm split}= \pm 20\%$,  is obtained when splitting the data in two zenith angle bands
(see Table \ref{table:HESSIISystematicErrors}). 

The energy range of the spectral fit is $\sim$10-110~GeV and was derived using MC simulations. 
Indeed, there is a non-negligible offset between the true energy and reconstructed energy scales due to
the large bias and dispersion in the reconstructed energies near the threshold (see Fig.~\ref{fig:MonoRecoAreaEnergydispersion}, right panel).
Of the total of 15835 excess events,  14415 are retained for the spectral fit with the nominal threshold (Cuts~I); those with a 
reconstructed  energy  $ E_{\rm rec} < E^{\rm safe}_{\rm rec}$ 
are  excluded. The number of events with a
true energy $E_{\rm true}< $  20~GeV participating to the spectral fit
under the power-law hypothesis (see Appendix \ref{sec:CT5Validation}) amounts up to  $\sim$5800 events, out of which 15\% lie below 10~GeV (i.e. 6\% of the total). 
The threshold energy for the P2 spectrum as measured by \hessII{}-CT5 was consequently chosen to be 10~GeV. This is further strengthened by  
the fact that the spectral index $\Gamma_{\rm HESS}^{\rm CI} = 4.06 \pm 0.16$ best compares with that of the \fermi-LAT $\Gamma_{\rm LAT} = 4.10 \pm 0.08$, which was obtained 
for a threshold  $E_{\rm Thresh}^{\rm LAT}=$~10~GeV. 

At the other end of the spectrum, the last significant bin covers the energy range, $E_{\rm rec} \in [92-110]$~GeV,  and entails 
912 excess events at a significance level of 3.3$\sigma$.
Owing to contamination from lower energy bins, the average true energy 
in this bin, under the power-law hypothesis,
is  $\langle E_{\rm true} \rangle = 82$~GeV, where RMS, 
$\sigma_{\langle E_{\rm true} \rangle}=29$~GeV, and a portion, $\rho_{>80\rm GeV}=33$\%, of events are predicted to lie above 80~GeV.     
When assuming the ECPL  or the LPB models derived from the \fermi-LAT data (see section \ref{sec:FermiSpectra}, above), 
these figures do not change significantly, i.e.  $\langle E_{\rm true} \rangle = 80$~GeV, $\sigma_{\langle E_{\rm true} \rangle}=28$~GeV and $\rho_{>80\rm GeV}=30$\%.  
As a result, the spectrum entails events with true energies ranging from below 10~GeV, up to  
$\langle E_{\rm true} \rangle + \sigma_{\langle E_{\rm true} \rangle}  \sim 110$~GeV. 

To test for curvature in the P2 spectrum, first, the LPB model was tested against the power-law hypothesis but 
resulted in unstable fits.  
Alternatively, the power law  was fitted to events selected using Cuts~II, i.e. above an approximately two times higher energy threshold 
(see section \ref{sec:HessIIEventReconstruction} and Appendix \ref{sec:HESSIISystematicErrors} for details).  
The spectral index obtained, i.e. $\Gamma_{\rm HESS}^{\rm CII}$ = $5.05\pm0.25,$ is significantly larger than the best-fit value found with 
the lower threshold analysis above.  Given that the two measurements share partially the same data, and thereby are correlated, 
the significance level of the deviation between the two indices exceeds 3$\sigma$. 
\begin{figure*}[htb]
\centering
\includegraphics[width=12.0cm]{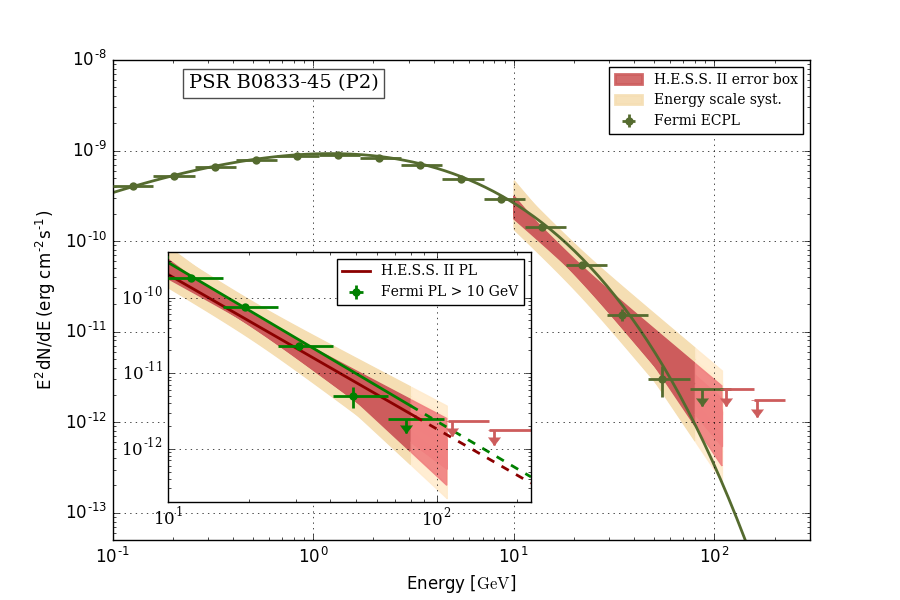}
\caption{Spectral energy distribution of the second peak, P2, of the Vela pulsar. 
The fit to the \fermi-LAT data above 100~MeV is shown as a dark green curve and points in the main frame.
Red indicates the error box in the 10-110~GeV energy range for the power-law fits to \hessII{} data.   
It consists of the union of $1\sigma$ statistical uncertainty confidence intervals obtained through the systematic error investigation procedure, including 
results from both analysis configurations, Cuts~I and II (see Appendix \ref{sec:HESSIISystematicErrors}).
The lighter colour above 80~GeV corresponds to the range
where the significance of detection is difficult to assess precisely (see text).
The error box has been extended to include a $\pm 10\%$ systematic error on the absolute energy scale. 
An upper limit of 8\% on the relative shift in the absolute energy scales of the two instruments
can be inferred based on the excellent agreement between the indices obtained under the power-law hypothesis (PWL). The inset shows the PWL fit 
to the \fermi-LAT data in solid green, and the \hessII{} best-fit power law as a dark red line.
Both are extrapolated above 80~GeV by dashed lines only to ease readability.      
The favoured hypothesis is a power law with an exponential cut-off (ECPL; see Table \ref{tab:FermiSpectra}). 
All upper limits are derived as 99.7\% confidence intervals.} 
\label{Fig:Spectrum}
\end{figure*}

The spectral fit results for P2 are presented as a confidence region in Fig.~\ref{Fig:Spectrum}, in the energy range 10-110~GeV 
where the region above 80~GeV 
is distinguished by a lighter colour. 
In order to take into account the variation of the spectral index with energy, the confidence region consists of 
the union of $1\sigma$ error boxes obtained under the power-law hypothesis for the two energy thresholds, including 
systematic  errors, as discussed in Appendix \ref{sec:HESSIISystematicErrors}.
Above 110~GeV, the 99.7\%  confidence level upper limits are derived in two bins with energy ranges of [110-157] and [157-225]~GeV.

Although the significance of LW2 is low in the \hessII{} data (see Table \ref{tab:HessVelaStatistics}), 
its spectrum was tentatively derived using the PWL model with both analysis configurations (the lack of statistics forbids any other meaningful model test). 
In contrast to the behaviour observed in the case of P2,  the photon indices obtained for LW2 with 
the nominal threshold cuts, $\Gamma_{\rm CI} =3.72\pm0.51$, and the higher threshold analysis,  
$\Gamma_{\rm CII} =3.48\pm0.21 $,
do not show any significant variation.
The energy bin [92-110]~GeV displays an excess of 343 excess at 2.5$\sigma$, but the 
next bin, [110-157] GeV, 
shows also an excess of 251 events at a significance level of 2.1$\sigma$.
Assuming the spectral models derived from the LAT data, the average true energy in the combined bin, [92-157]~GeV, varies 
 from $\langle E_{\rm true} \rangle \sim$110~GeV for a simple power law with $\Gamma_{20~\rm GeV} = 2.80$, 
to $\langle E_{\rm true} \rangle \sim$130~GeV for the BPL form.
The combined excess (594 events at 3.3$\sigma$) represents 30\% of the total excess measured from LW2 
with Cuts~II (see Table \ref{tab:HessVelaStatistics}); for P2 this ratio is only $\sim$13\%.   

Along with the spectral fit results, these numbers 
point to the harder nature of the LW2 spectrum, as compared to P2, in the tens of GeV range,
and thereby support the indications found in the \fermi-LAT data in 
sections~\ref{sec:FermiLightCurve} and \ref{sec:FermiSpectra}, and in the analysis of the \hessII{} light curve in \ref{sec:HessIILightcurve}.
\section{Discussion}
\label{sec:Discussion}
\subsection{Relative energy scale offset between \fermi-LAT and \hessII{}}
\label{sec:EnergyScalesDiscussion}
The Vela signal is a unique occasion to compare the energy scale of a ground-based telescope such as \hessII{} with 
that of the beam-calibrated LAT instrument. Indeed, the pulsed nature of the emission enables one
to extract the on- and off-source events from the same
portion of the field of view, thereby eliminating a significant source of systematic errors that arise from variations of
acceptance as a function of direction in the sky and/or position in the camera. 
The power-law index for P2 as measured by \hessII{} is in excellent agreement with that derived with the \fermi-LAT above 10~GeV, where  
$\Gamma_{\rm HESS}^{\rm CI} = 4.06 \pm 0.16$ and $\Gamma_{\rm LAT}^{10\,\rm GeV} = 4.10 \pm 0.08$. While the flux normalizations 
show a deviation, their  ratio, $\Phi_0^{\rm HESS}/\Phi_0^{\rm LAT} = 0.76 \pm 0.06^{\rm stat} \pm0.21^{\rm sys}$
(see Fig.~\ref{Fig:Contours} and the inset in Fig. \ref{Fig:Spectrum}), remains compatible with unity,
given the systematic uncertainties (see Appendix \ref{sec:HESSIISystematicErrors} and \ref{sec:FermiSystematicErrors}).
It is noticeable that $\Phi_0^{\rm HESS}/\Phi_0^{\rm LAT}$ is stable with respect to variations of $E_{\rm Thresh}^{\rm LAT}$ from 8 to 20~GeV. 

If the deviation in flux is assumed to be only due to a difference in absolute energy scales, 
a relative offset,  $\Delta_{E}^{\rm scale} =(E_{\rm LAT}-E_{\rm HESS})/E_{\rm LAT} \leq$8\%, can be inferred 
between the two instruments. However, as other systematic effects that can bias the effective area (e.g. the uncertainties on event 
reconstruction and/or selection efficiencies) are not 
excluded, this value has to be considered as a conservative upper bound.   
With regard to the absolute energy measurements, this value of $\Delta_{E}^{\rm scale}$ is well contained within the uncertainty range of 
$\pm 10$\% usually quoted for IACTs (e.g. \cite{hess:Crab,MayerHorns2010}). Additionally,
the systematic error estimated for the absolute scale of the \fermi-LAT, $+2\%$/$-5\%$~\citep{2012ApJS..203....4A}, 
should be taken into account.

\subsection{Evolution of the light curve as a function of energy}
\label{sec:P2Morphology}
The fit of a three-component function (two asymmetric Lorentzian and a log-normal function) to the \fermi-LAT data in different energy bands confirms the main characteristics of the Vela pulsar light curve 
that were previously
revealed with {\it COS B} \citep{VelaCosBKanbach80}, elaborated with {\it EGRET} \citep{VelaEgretKanbach94} and {\it  AGILE} \citep{AgilePulsars2009}, and then 
subsequently measured up to 20~GeV with the \fermi-LAT \citep{fermifirstvela, Abdo2010Vela}.
Namely, we observe, with increasing energy: (i) a decrease in the P1/P2 ratio; (ii) a sharpening of the outer edges of 
both peaks; (iii) a continuous decrease of the inner width of P2, while that of P1 attains a maximum in the 3-10~GeV band before decreasing; 
(iv) no measurable change in the P1 and P2 positions; and (v) a shift to later phases of P3.  
As mentioned in section \ref{sec:P2properties}, the asymmetric Lorentzian functional form does not describe fully the data, and  
the analysis of the light curve with a KDE, i.e. with no strong a priori assumptions on its form, results in the following two differences: 
(vi) an offset of few milli-periods is obtained between the fitted positions of P1 and P2 and their maxima and (vii) the maximum of P1 shifts to later phases with increasing energy. 
   
While \hessII{} data below 33~GeV confirm the evolution with energy of the P1/P2 ratio and  P2  noted in (i), (iv), and (vi) above,
in the tens of GeV energy range, 
a qualitative change of P2 is found. A shift to later phases -- by $\sim$5 milli-periods -- of its fitted position is observed, 
which can be attributed to an extreme sharpening of its trailing edge, together 
with the possible onset of a new component at 3.4$\sigma$ significance level.
These, combined with the hardening of LW2 above 50~GeV, could be at the origin of the widening of the leading edge of P2. 
 
The behaviour of the amplitude ratio of P1 and P2 as a function of energy is in line with results obtained for some other bright
\fermi-detected pulsars \citep[e.g.][]{Abdo2010Crab,Abdo2010Geminga, Abdo2010EGRET}. 
The physical processes responsible
for this trend are not known, although
within the context of curvature radiation in the
radiation-reaction limit, this trend may be pointing to relatively
weaker electric fields and/or smaller curvature radii of magnetic
field lines in the magnetospheric regions where P1 originates,
compared to the P2 regions. \citet{Bednarek2012} proposed a light
cylinder gap model in which the leading and trailing magnetic field
lines have different radii of curvature, leading to different spectral
cut-offs for the two main peaks. \citet{Hirotani2014}, however, showed
that this model is based on the erroneous assumption of divergence of the
Goldreich-Julian charge density at the light cylinder.  Furthermore,
in the popular caustic models, the two pulses are not formed by
leading and trailing field lines, but rather by caustics (where
photons accumulate in phase) formed by either trailing or overlapping
magnetic field lines \citep{Dyks2004}. The P1/P2 trend has furthermore
not been reproduced (or predicted) by recent 3D numerical
magnetospheric gap models of  $\gamma$-ray pulsars
\citep[e.g.][]{Wang2011}. On the other hand, \citet{Brambilla2015}
introduced a pulsar model with the accelerating electric field
operating outside the light cylinder and found that in about half of
their predicted light curves, a larger energy cut-off value was
produced in P2 compared to P1 because of a larger azimuthally dependent
electric field in that emission region.

A decrease in pulse width with increasing energy was also seen for the Crab pulsar
\citep{Abdo2010Crab, Aliu2011,Aleksic2012}. This phenomenon may point
to the fact that the particles responsible for high-energy
emission are confined to a smaller region embedded within the
 $\gamma$-ray emitting zone, corresponding to general expectations of
magnetospheric gap models, where the accelerating electric field is
zero at the gap boundaries, but peaks in its centre
\cite[e.g.][]{Muslimov2003,Wang2010}.
Wind models, in turn, naturally explain a pulse width that decreases with increasing energy, where the
high-energy pulsed emission is due to Doppler-boosted synchrotron
radiation  by relativistic electrons powered by magnetic field
line reconnections in the wind current sheet \citep[e.g.
][]{Arka2013,Mochol2015}.

\subsection{Spectral shape of P2 in 10-80~GeV range}
\label{sec:P2Spectralshape}
Although a very good agreement is found between the power-law                                               
indices derived for the \fermi-LAT data above 10~GeV and for \hessII{} with the Cuts~I analysis 
configuration, the power law is not the favoured model for the spectrum of P2 at these energies.
Indeed, the curvature measured with the LAT at a significance level of $3.3\sigma$ assuming the LPB model, which has been shown to be robust against systematic
uncertainties in Appendix \ref{sec:FermiSystematicErrors}, is confirmed by   
the variation observed in the spectral index as a function of the analysis threshold energy for both instruments, in a 
consistent manner: 
$\Delta_{\Gamma_{\rm LAT}}^{10-20 \rm \,GeV}={+0.70}\pm0.31$  and  $\Delta_{\Gamma_{\rm HESS}}^{\rm CI-CII}={+0.90}\pm0.30$. 
The fitted values of $\Gamma_{\rm HESS}$ and  $\Gamma_{\rm LAT}$ correspond hence to the average slope of a curved spectrum (i.e. in a log-log plot) 
above a given threshold, rather than corresponding to the index of a power law. The signal from P2 detected by \hessII{}  
consists therefore of the same spectral component as that 
of the \fermi-LAT data above 100~MeV, and thereby confirms its sub-exponential cut-off (ECPL) form. 

\citet{Abdo2010Vela} and \citet{Leung14} already showed 
the ECPL nature of the phase-averaged spectrum of the Vela pulsar.
In addition to P2, our analysis of the eight-year LAT data shows a clear preference
for the ECPL model for the phase-resolved spectra of P1, LW2, and P3. 
The traditional outer gap models with a single value of injected
current  may have difficulty reproducing this spectrum
invoking the usual curvature radiation component, given its high-energy sub-exponential
form. \citet{Leung14} therefore proposed a `superposition of
stationary outer gap states' to match their \fermi-LAT
spectrum. However, this model implies an increase in pulse width with
energy (cf.\ their section~3), contrary to what is seen. 

\subsection{Hints of pulsed emission $>$100~GeV from the Vela PSR}
\label{sec:PulsedemissionAbove80GeV}
Estimation based on the energy migration matrix shows that under the ECPL hypothesis for P2, with best-fit parameters of the LAT spectrum, 
\hessII{} data contains more than 2000
events above 50~GeV in the [0.5, 0.6] phase range, and that the emission extends at 
least up to an energy of 80~GeV. This corresponds to the average energy of the highest energy significant 
bin in the data ([92-110]~GeV) using Cuts~I; 912 events are at a significance level of $3.3\sigma$ of which a portion, $\rho_{>80\rm GeV}=$30\%, 
is predicted to lie above 80~GeV.  With the tighter Cuts~II, the excess in this bin drops to 620 events at  $3.0\sigma$, while the next bin ([110-157]~GeV),
which has an estimated $\langle E_{\rm true} \rangle \sim$110~GeV, 
shows an excess of 334 events at a level of $1.8\sigma$ only.

The 96 months \fermi-LAT data set contains, in turn, 31 photons above 50~GeV, of which 7 lie in the P2 phase range and only two photons exhibit an
energy above 80~GeV, i.e. 93.7 and 206.3~GeV. The P2 signal above 90~GeV found by \citet{Leung14} at a significance level of 3.3$\sigma$ consisted of 
these two photons, detected within the first 62 months of data, and with the \texttt{P7REP} processing.  
Here, with the \texttt{P8} processing, the source probability (i.e. the estimated probability for a photon to originate from the pulsar) of 
the 206.3~GeV event has dropped from $P_{\rm PSR}^{\texttt{P7REP}}$=92.2\% 
to $P_{\rm PSR}^{\texttt{P8}}$=1.8\%, mainly due to
its larger reconstructed angular distance to the source ($\Delta\theta_{\texttt{P8}}$=$0.33^\circ$, as compared 
to $\Delta\theta_{\texttt{P7REP}}$=$0.092^\circ$; see Table~\ref{table:Fermi80GeVEvents}).
Still, a phase-resolved analysis in the P2 range above 80~GeV results in a test statistic value, TS=9.8, i.e. a significance of  $\sim$3$\sigma$.
Hence, both instruments give independent evidence for a weak signal in the P2 phase range above 80~GeV and at a significance level of $\sim$3$\sigma$ each.

The leading wing of P2, LW2, shows different behaviour. As discussed in section \ref{sec:FermiSpectra}, the log-parabola (LPB) fit to \fermi-LAT data 
results in a convex curve, 
suggesting a hardening of its spectrum with increasing energy. This is confirmed by a  BPL fit, which yields 
a break energy $E_b=50.2 \pm9.5$~GeV, and indices $\Gamma_{1\,\rm} = 4.37 \pm 0.24 $ and  
$\Gamma_{2\,\rm} = 1.37 \pm 0.64 $.  
Although the two indices differ significantly from each other, a likelihood-ratio test shows that the BPL is favoured only at  $S_{\rm BPL} =2.3$$\sigma$.  
At the same time, a power-law fit above 80~GeV yields  an index,
$\Gamma_{80~\rm GeV} = 1.80 \pm 1.1$, which is well compatible with $\Gamma_{2\,\rm}$, and a test statistic  value,  $\rm TS =$ 16.8. 
This corresponds to an evidence for a signal above 80~GeV at a significance level of $\sim 4$$\sigma$.   
Investigation at the event level shows indeed that five out of the fifteen 
events selected with an energy $>80$~GeV (and within a radius of $\theta_{\rm max}=0.8^\circ$; see section \ref{sec:FermiAnalysis}) 
lie in the phase range corresponding to LW2, [0.45-0.5], i.e. 5\% of the full rotation period, or a chance probability of $<0.001$ ($>3\sigma$).   
In addition, four of these events display a high probability of originating from the pulsar, $P_{\rm PSR}$, ranging from 77\% to $>99$\% 
(see Table \ref{table:Fermi80GeVEvents}). It is remarkable that all events except one have an energy exceeding 100~GeV, of which
the highest energy photon exhibits 377~GeV together with the highest source probability, $P_{\rm PSR}>$99.3\%. 
We note that none of these events display any peculiarity,
regarding  reconstructed angles in the instrument, conversion types, or zenith angles.
We note also that two (four) of the other $>80$~GeV events  are within the P1 phase interval (P3, respectively), and that none
are detected in the background interval (see Table~\ref{table:Fermi80GeVEvents}). 

The hardness of the LW2 spectrum in the tens of GeV range is further supported by the 
\hessII{} data: (i) the analysis of the \hessII{} light curve in the LW2 phase range (section \ref{sec:HessIILightcurve}) resulted in 
a higher significance (4.5$\sigma$) with the higher threshold configuration, Cuts~II, as compared to the nominal threshold analysis, Cuts~I (3.4$\sigma$); 
(ii) contrary to the steepening observed for P2 with the latter configuration, the photon index obtained for LW2, $\Gamma_{\rm HESS}^{\rm CII} =3.48\pm0.21 $,  
does not show any significant variation as compared to $\Gamma_{\rm HESS}^{\rm CI} = 3.72\pm0.51 $, which is derived with the  nominal threshold analysis, 
Cuts~I (section \ref{sec:HessIISpectra}); and (iii) although the LW2/P2 flux ratio at 10~GeV is of only $\sim$12\% (see Table \ref{tab:FermiSpectra}),
the highest energy bins of the LW2 spectrum display excess counts and significance levels comparable to those obtained for P2, in which a total of 
594 events in the [110-157] GeV range, corresponding to  $\langle E_{\rm true} \rangle \gtrsim$100~GeV, were detected at a significance level of 3.3$\sigma$. 
There are therefore converging indications from both instruments that the emission from 
LW2 is harder than that from P2 and that it extends beyond 100~GeV, i.e. to the very high-energy (VHE) range, in contrast to the weak signal from P2 itself.

The Crab pulsar is the only pulsar known to emit at energies beyond 100~GeV \citep{Aliu2011,Aleksic2012}. 
In the case of Geminga, the second brightest $\gamma$-ray pulsar in the GeV sky, only upper limits have been
derived above $100$~GeV so far \citep{Aliu2015,Ahnen2016}. \citet{McCann15} performed a stacking analysis involving 115
\fermi{} pulsars (excluding the Crab) and did not find any
significant emission above $50$~GeV.
The VHE emission is detected from both peaks of the Crab pulsar and also 
from the bridge \citep{Aleksic2014}. More recently, \grs{} reaching  1.5~TeV \citep{Ansoldi2016} were reported from the 
second peak of the pulsar. 
There have been a number of attempts to explain \citep[e.g.][]{Aleksic2012,coldwindaharonian2012} and predict the VHE emission by pulsars
\citep[e.g.][]{Muslimov2003,Muslimov2004,Du12}.
No significant VHE emission is expected so far for current IACTs from the Vela pulsar,  \citep[e.g.][]{Harding2015,Mochol2015}.
It should be noted that all VHE components of the Crab pulsar seem to connect smoothly with their lower energy counterparts, 
i.e. a simple  power law describes well the data above 10~GeV.     
This would not be the case for LW2, given that if the evidence for its hardening above 50~GeV is confirmed, it should correspond to the rise of
a second and new component.

\section{Summary}
\label{sec:Summary}
Pulsed \gr{} emission from the Vela pulsar was detected at high significance with the largest telescope of the \hessII{}~array, CT5. This telescope 
was added as an upgrade to the initial four-telescope set-up in 2012 in view of lowering its detection energy threshold down to a few tens of GeV.
Data from 40.3~h of observations have been analysed in monoscopic mode, through a 
reconstruction pipeline specifically designed to achieve a large effective 
area at the lowest energies possible.     
Data from eight years of \fermi-LAT observations were analysed in parallel and used as input to MC simulations of the overall detection chain,  
and subsequently used for comparison to the \hessII{} results. 
An excellent agreement was obtained and made it possible to validate the response model and the analysis pipeline of CT5 down to 
the sub-20~GeV range  with reasonable systematic uncertainties. An
upper limit on the relative offset in the energy scales of the two instruments, $\Delta_{E}^{\rm scale} =(E_{\rm LAT}-E_{\rm HESS})/E_{\rm LAT} \leq$8\%,
could be consequently derived.

Measurement of spectra extending to the sub-20~GeV domain is unprecedented
in ground-based $\gamma$-ray astronomy. We note, however, that 
the pulsed nature of the signal plays a major role here, as it enables                                                                                          
extraction of the on- and off-source events from the same                                                                                                                    
portion of the field of view, thereby eliminating a major part of systematic effects that arise from variations of
acceptance as a function of direction in the sky and/or position in the camera.

The study of the Vela pulsar light curve and its energy dependence with the \fermi-LAT confirmed its main and previously known characteristics up to 20~GeV.
Beyond this energy, a shift to later phases was found for P2 in the \hessII{} light curve, possibly owing 
to a change of morphology and the onset of a new component at a confidence level of 3.4$\sigma$. 
The measurement of the P2 spectrum above different energy thresholds with the two instruments demonstrated its curved form in the 10 to 80~GeV range; there was 
only weak evidence for a signal above 100~GeV.
In contrast, the leading wing of P2 was shown to possibly exhibit a hard component setting in above $\sim$50~GeV, 
with hints of extension beyond 100~GeV, namely, a 4$\sigma$ signal above 80~GeV including 4 events with energies $>$100~GeV in the LAT data
and 594 events above 100~GeV in the CT5 data at 3.3$\sigma$.    

\begin{acknowledgements}
The support of the Namibian authorities and of the University of Namibia in facilitating 
the construction and operation of H.E.S.S. is gratefully acknowledged, as is the support 
by the German Ministry for Education and Research (BMBF), the Max Planck Society, the 
German Research Foundation (DFG), the Helmholtz Association, the Alexander von Humboldt Foundation, 
the French Ministry of Higher Education, Research and Innovation, the Centre National de la 
Recherche Scientifique (CNRS/IN2P3 and CNRS/INSU), the Commissariat à l’énergie atomique 
et aux énergies alternatives (CEA), the U.K. Science and Technology Facilities Council (STFC), 
the Knut and Alice Wallenberg Foundation, the National Science Centre, Poland grant no. 2016/22/M/ST9/00382, 
the South African Department of Science and Technology and National Research Foundation, the 
University of Namibia, the National Commission on Research, Science \& Technology of Namibia (NCRST), 
the Austrian Federal Ministry of Education, Science and Research and the Austrian Science Fund (FWF), 
the Australian Research Council (ARC), the Japan Society for the Promotion of Science and by the 
University of Amsterdam. We appreciate the excellent work of the technical support staff in Berlin, 
Zeuthen, Heidelberg, Palaiseau, Paris, Saclay, Tübingen, and in Namibia in the construction and 
operation of the equipment. This work benefitted from services provided by the H.E.S.S. 
Virtual Organisation, supported by the national resource providers of the EGI Federation. 

The Parkes radio telescope is part of the Australia Telescope, which is funded by the Commonwealth Government for operation as a National Facility managed by CSIRO.
Work at NRL is supported by NASA.

\end{acknowledgements}
\raggedbottom 
\bibliographystyle{aa}
\bibliography{32153_refs_final}

\appendix{}
\section{Validation of \hessII{} CT5 response model and  analysis pipeline}
\label{sec:CT5Validation}
\begin{figure*}[tb]
\centering
\includegraphics[width=8.4cm]{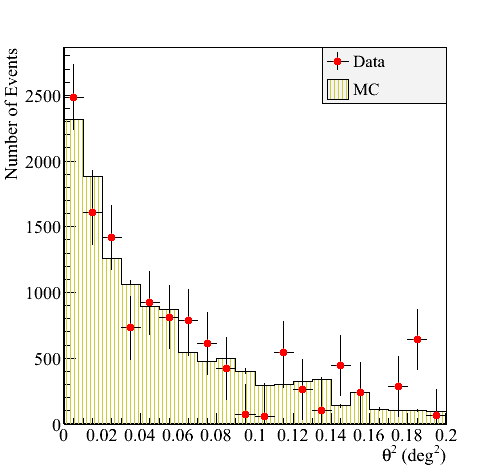}
\includegraphics[width=8.4cm]{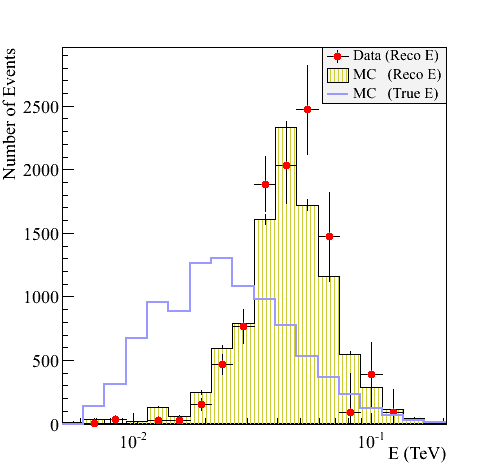}
\caption{Distribution of the square of the angle between the source position and
event direction ({\it left}), and distribution of the reconstructed energy  ({\it right}) for data (excess events) and MC simulations. 
The latter have been weighted such as to represent the power law fitted to the \fermi-LAT data with index $\Gamma = 4.1$ and scaled by a factor $\eta =0.84$ (see text).
The light blue histogram on the right panel is the corresponding distribution for the generated true energy, $E_{\rm true}$, for MC events passing the analysis cuts. 
This distribution has an average energy of 31~GeV and peaks at $\sim$20~GeV: $\sim 40$\% of events lie below the latter energy,  
34\%  are in the 10-20~GeV range, and 6\% have an energy below 10~GeV.
\label{Fig:Theta2AndEnergyDistribution}}
\end{figure*}

 Relatively good knowledge of the source spectrum above 10~GeV,
thanks to the analysis of \fermi-LAT data,
enables us to use the Vela pulsar emission as a
test beam to check the validity of the overall analysis pipeline, i.e. 
the instrument response model obtained through
MC simulations, and the methods for event direction and energy reconstruction.  
Data set I of the commissioning period was initially used for such a study, where 
the power law obtained from 
\fermi-LAT data above  $E_{\rm Thresh}^{\rm LAT}=$10~GeV on P2 (with photon index $\Gamma_{\rm LAT} = 4.10 \pm0.08$, and normalization, 
$\Phi_0=40.3 \pm 1.8 \times 10^{-9}$ {TeV}$^{-1}$cm$^{-2}$s$^{-1}$, at a reference energy, $E_0=25$ GeV; see  section~\ref{sec:FermiSpectra})
 was used to produce the MC-expected signal and 
corresponding low-level parameters distributions, which were subsequently compared to measurements.   
The expected number of $\gamma$-ray events, $N_{\rm MC}= 11697 \pm 675$, for which the error is evaluated using  
the statistical uncertainty on the \fermi-LAT flux normalization (see section \ref{sec:FermiLightCurve}),  
compares well to the measured excess, $N_{\rm HESS}= 9789 \pm 789$ (see section \ref{sec:HessIILightcurve}).  
Although the deviation of $1908 \pm 1038$ events, corresponding to a ratio $\eta = N_{\rm HESS}/N_{\rm MC}=0.84 \pm 0.08$, 
is of low statistical significance ($\lesssim 2\sigma$),
it can point to systematic errors in the CT5 effective area calculation 
and/or an offset between the energy scales of the two instruments (see Appendix \ref{sec:SystematicErrors}, below), 
and/or deviations of the intrinsic source spectrum from the assumed model.
Regarding this point,  varying the power-law model index, $\Gamma_{\rm LAT}$, from 3.86 to 4.55 (corresponding to variations of $E_{\rm Thresh}^{\rm LAT}$ 
from 8 to 15~GeV; see Appendix \ref{sec:FermiSystematicErrors}), implies 
values for $\eta$  ranging from 0.80 to 0.91, respectively. Alternatively, using the power law with an exponential cut-off obtained from the 
fit to the \fermi-LAT data above 100~MeV (ECPL; see Table \ref{tab:FermiSpectra}), yields $\eta=0.91\pm 0.09$.

Measurements were compared further with MC simulations using low-level reconstructed parameters.
Fig.~\ref{Fig:Theta2AndEnergyDistribution} shows the distribution of the square of the angle between the source position and
event direction, $\theta^2$, and of the reconstructed energy, $E_{\rm rec}$, for excess events and for MC simulations,
after scaling with $\eta=0.84$. 
The agreement between expected and experimental
distributions validates the overall analysis chain and the MC
model of the instrument down to its threshold energy. 
The true-energy distribution of MC events passing the analysis cuts is shown in blue on the right panel of Fig.~\ref{Fig:Theta2AndEnergyDistribution}.
The distribution peaks at 20~GeV with an average energy of 31~GeV. It is noteworthy that  $\sim 40$\% of events lie below the peak, out of which 15\% have 
an energy < 10 GeV. These figures do not change when considering events above $E^{\rm safe}_{\rm rec}$ for the spectral derivation (see section \ref{sec:HESSIISystematicErrors}). 
A significant overlap in energy range can thus be inferred
between \hessII{} in monoscopic mode and \fermi-LAT.

\section{Systematic errors on spectral parameters}
\label{sec:SystematicErrors}

\subsection{\hessII{} systematic errors}
\label{sec:HESSIISystematicErrors}
Spectral measurements close to the trigger threshold of IACTs are challenging. 
One important limitation is related to the systematic uncertainties on the effective area   
near the threshold, mainly due to the lack of precise knowledge on the atmospheric transparency, regarding both its absolute value 
and its variations. While the former uncertainty is typically considered to lie within $\pm 10$\%, and imply an error of the same magnitude
on the absolute energy scale of the IACTs, one seeks to limit the latter by monitoring the atmospheric conditions during the data taking and by
selecting data, accordingly. This process entails a tolerance (in terms of the degree of severity of the quality cuts) and leads in turn to
systematic errors, especially close to the detection threshold. Fluctuations of the night sky background, as well as residual instrumental effects 
that are difficult to model precisely, add to the uncertainties on the effective area close to the threshold. This implies possible biases on
the measurement of a given source spectrum, both for the absolute flux and the spectral index (e.g. for a power law).
To limit these errors, a safe energy threshold cut is applied on the reconstructed energy, $ E_{\rm rec} < E^{\rm safe}_{\rm rec}\simeq 20$~GeV. The 
cut value is determined using
MC simulations of the spectral fitting process with manual introduction of errors in the IRFs,  for example by scaling the effective area 
by error functions inferred from the variance of low-level parameters of background events (trigger rate and image charge distributions), recorded 
in similar conditions. 
Given that there is some degree of arbitrariness in this process,
possible systematic effects due to the particular choice of the scaling function, and hence of $E^{\rm safe}_{\rm rec}$, were investigated by testing different values 
of the latter (ranging from 20 to 45~GeV), including the case in which no such cut was applied.  
These tests have shown that systematic variations do not exceed the amplitude of statistical errors, i.e. $\delta_{\Phi_0}^{\rm E_{\rm safe}}=\pm 10\%$ and
$\delta_{\Gamma}^{\rm E_{\rm safe}}$=$-0.2$ to $+0.3$, when  $E^{\rm safe}_{\rm rec}$ is varied from no cut up to 45~GeV.
The variation of the index as a function of $E^{\rm safe}_{\rm rec}$ could be assumed to result, at least partly, from the curvature of the P2 spectrum (as measured with 
\fermi-LAT, see section \ref{sec:FermiSpectra} and Appendix \ref{sec:FermiSystematicErrors}, below). However, because of the large bias and dispersion in 
the reconstructed energies near the threshold (leading  
to large migration of events from lower to higher energies), the cut on  $E^{\rm safe}_{\rm rec}$ does not result in a sharp rise of the energy threshold; hence such an assumption has to be checked.  

An alternative method consisting of applying a tighter cut on image intensities,  $Q_{\rm min}=60$ p.e., that is two times larger
than the standard cut (Cuts~II, see section \ref{sec:HessIIEventReconstruction}), is 
better suited for this purpose. 
Given that image intensities are roughly proportional to the energy of \grs{}, this implies a 
higher threshold,  $E_{\rm true}^{60\, \rm p.e.} \sim 2 \times  E_{\rm true}^{30\, \rm p.e.} $.    
The spectral index obtained with this configuration, $\Gamma=5.05\pm0.25$,  deviates strongly from the best-fit value of the standard 
analysis, $\Gamma=4.06\pm0.16$, with a significance level in excess of 3$\sigma$ ; we note that the errors of the two measurements, which share partially the same data,  are correlated.  
The variation observed in the \hessII{} data, $\Delta_{\Gamma}^{\rm {\tiny HESS}}$=$+0.90 \pm 0.3$, is consistent with that obtained 
with \fermi-LAT, $\Delta_{\Gamma}^{\rm {\tiny LAT}}$=${+0.70}\pm0.30$,  when its analysis threshold,  $E_{\rm Thresh}^{\rm {\tiny LAT}}$,  is increased from 10 to 20~GeV (see section \ref{sec:FermiSystematicErrors}).
One can hence conclude with confidence that the smaller amplitude of the above-mentioned variation of the index with $E^{\rm safe}_{\rm rec}$, $\delta_{\Gamma}^{\rm E_{\rm safe}}$=$-0.2$ to $+0.3$, should be mainly due to the curvature in the spectrum, rather than resulting from a systematic effect. 

An estimate of the magnitude of possible systematic effects due to the uncertainty in the absolute value of the atmospheric 
transparency was obtained using a  different extinction model.
The test of a model with a 45\% larger aerosol optical 
depth at 400~nm (from 10~km to the site altitude) resulted in a 
larger flux normalization, $\delta_{\Phi_0}^{\rm Atm}\simeq +10\%$, and a small change,  $\delta_{\Gamma}^{\rm Atm}$=$+0.08$, of the spectral index. 
We note, however, that the comparison of the \fermi-LAT and \hessII{}-CT5 power-law fits makes it possible to constrain significantly this source 
of systematic error (see section \ref{sec:EnergyScalesDiscussion}). 

\begin{table}[h]
\center
\begin{footnotesize}
  \caption{Investigation of systematic errors on the power-law best-fit values for the P2 spectrum in the range $\sim$10-110~GeV.
Results are shown for different data sets (I and II) and for two zenith angle bands; the overall data set
uses\ two atmospheric extinction models. 
\label {table:HESSIISystematicErrors}}
\centering
\begin{minipage}{\textwidth}
  \begin{tabular}{ccccc} 
  \hline
Data set & Extinction & Zenith &$\Phi_0$\footnote{In units of $10^{-9}$TeV$^{-1}$cm$^{-2}$ s$^{-1}$ at $E_0=25 $GeV.} & $\Gamma$ \\
\hline
\hline
I+II     & standard    & all  & $30.6 \pm 1.9 $  & $4.06\pm 0.16$\\
I+II     & alternative & all  & $34.4 \pm 4.5 $  & $4.14\pm 0.20$\\
\hline
I        & standard    & all  & $30.7 \pm 3.1 $ &  $4.14 \pm 0.28$\\
II       & standard    & all  & $28.8 \pm 3.4 $ &  $3.94 \pm 0.28$\\
\hline
I+II     & standard    & $<23.7^{\circ}$& $37.1 \pm 3.1 $  & $3.96\pm 0.22$\\  
I+II     & standard    & $>23.7^{\circ}$& $25.4 \pm 2.6 $  & $4.03\pm 0.28$\\  
\hline
  \end{tabular}
\end{minipage}
\end{footnotesize}
\end{table}

Further investigation of systematic effects was made by splitting the data in several ways. Fitting 
data sets I and II separately shows results that are compatible with each other and with the overall data set 
within statistical uncertainties (see Table \ref{table:HESSIISystematicErrors}). Splitting the data in two zenith angle bands, however, i.e. below and above the median zenith angle of $23.7^\circ$, results in stable indices of $\delta_{\Gamma}^{\rm split}$=$-0.1$, 
but flux normalization deviations of $\delta_{\Phi_0}^{\rm split} \lesssim \pm 20\%$, which are larger than statistical errors. 

For the overall systematic error on the best-fit value of the spectral index
we retain, conservatively, $\delta_{\Gamma}^{\rm sys}$ = $\delta_{\Gamma}^{\rm E_{\rm safe}}$=~$-0.2$/$+0.3$.  
For the flux normalization,  the quadratic combination of $\delta_{\Phi_0}^{\rm split}$$\sim$$\pm 20\%$ 
and $\delta_{\Phi_0}^{\rm Atm}\simeq +10\%$, i.e.   $\delta_{\Phi_0}^{\rm sys}=-$20\%/$+$25\%, is used.  

These errors are independent of statistical fluctuations and have been added quadratically to the 1$\sigma$ statistical error 
contour of the overall data set in Fig.~\ref{Fig:Contours}.
Alternatively, the central error box in Fig.~\ref{Fig:Spectrum} represents the union of all $1\sigma$ statistical uncertainty confidence intervals 
obtained through the systematic error investigation procedure, including those computed for the higher threshold (Cuts~II) analysis results. In addition, the error box has been 
extended such as to include the uncertainty of $\pm10$\% on the absolute energy scale. 
\subsection{\fermi-LAT systematic errors}
\label{sec:FermiSystematicErrors}
The systematic errors on \fermi-LAT results were studied mainly in the overlapping energy range with \hessII{}, i.e. above 10~GeV. One of the main sources of error is the
uncertainty on the spectral models used for extended, bright, hard and nearby sources, i.e. the supernova remnants Vela~Jr and Puppis~A, the Vela~X pulsar wind nebula, 
and the normalization of the Galactic diffuse emission model.  
Two extreme cases were modelled in which the flux normalization and spectral index of the above-mentioned sources were pushed to~i) maximum flux normalization 
and hardest index and ii) minimum flux and softest index. These numbers were 
determined from statistical and systematic errors reported in the dedicated papers for Puppis~A \citep{FermiPuppisA2012} and  Vela~Jr \citep{FermiVelaJr2011}.  
For the Vela~X nebula, the best-fit flux normalization was obtained through a dedicated off-pulse analysis in the restricted phase range [0.8-1.0] and 
using the same BPL definition as \cite{FermiVelaX2013}. We note that the normalization derived in this work (e.g. at the break energy of 2.1~GeV)  
is a factor 1.5 lower than that found in \cite{FermiVelaX2013}, while being very close to that of 3FGL catalogue. 
This is somewhat expected because the 3FGL catalogue was used as source model, but was not available in \cite{FermiVelaX2013}.       
In the study of systematic effects in the same paper, the deviation of the best-fit Galactic diffuse intensity for nearby source-free regions was shown to be $\leqslant 6$\%.  
Its normalization was conservatively modified by $+10$\% and $-10$\% in the two cases i) and ii), respectively. 
Comparison  of the best-fit parameters for these extreme cases has shown that the spectral parameters for the two models used in section \ref{sec:FermiSpectra}, i.e. a power law  and a log parabola (LPB), have no deviations beyond the statistical errors; 
that is a systematic error of $\delta_{\Phi_O}^{\rm Model}=\pm 2$\% for the flux normalization and of $\delta_{\Gamma}^{\rm Model} =\pm 0.02$ for the index, which has no measurable effect on the curvature parameter of the LPB (becasue of  its large statistical error, $\beta = 0.7\pm 0.3$).
However, the confidence level for the LPB hypothesis, when tested against a simple power-law model, 
showed some variation, i.e. 3.1$\sigma$ to  3.5$\sigma$ for cases i) and ii), respectively.  
Checking the effect of each source model individually shows that the main contributor to these variations is the Galactic diffuse model normalization. 
In addition, the effect of the uncertainties on the effective area were studied using modified IRFs, which can simulate 
instrument model deviations from the real instrument.
The application of the effective area bracketing method\footnote{As recommended 
in \url{https://fermi.gsfc.nasa.gov/ssc/data/analysis/scitools/Aeff_Systematics.html}} resulted  in differences of spectral 
parameters from best-fit values that were smaller than the estimates given for a soft 
source in \cite{2012ApJS..203....4A}, i.e. $\delta_{\Phi_O}^{\rm IRF}$= $\pm 11$\% for the flux normalization and $\delta_{\Gamma}^{\rm IRF}$ =$\pm 0.1$ for the index.    
These values were consequently adopted as systematic errors for spectral measurements with the \fermi-LAT in this paper. 

The most important source of uncertainty, when comparing \hessII{} and  \fermi-LAT results  and assessing the significance of the LPB model as compared to the power-law hypothesis, 
is the LAT analysis energy threshold, $E_{\rm Thresh}^{\rm {\tiny LAT}}$, which was
initially set to 10~GeV. Testing  $E_{\rm Thresh}^{\rm {\tiny LAT}}$ values of 8, 12, 15, and 20~GeV,  
resulted in stable flux estimates, while the spectral index showed variation from $\delta_{\Gamma}^{\rm {\tiny Thresh}}$=$ {-0.25}\pm0.05$ at 8~GeV, to ${+0.45}\pm0.17$ at 15~GeV, 
and ${+0.70}\pm0.30$ for a threshold of 20~GeV. This behaviour is expected as well, given the possibly curved nature of the 
P2 spectrum at least up to $\sim 50$~GeV (see section \ref{sec:FermiSpectra} and Fig.~\ref{Fig:Contours}). The
significance of the LPB hypothesis also depends on the chosen threshold, 
varying from a very significant level,  $S_{LPB}$~=~$+7.3\sigma$ for $E_{\rm Thresh}^{\rm {\tiny LAT}}$~=~8~GeV, down to only $1.9\sigma$ 
at 12~GeV, and falling below $1\sigma$ above 15~GeV.

It is clear that the above-mentioned variation of $\Gamma_{\rm LAT}$ as a function of energy does not constitute a systematic error on 
the \fermi-LAT measurements themselves, but should be taken into account when it comes to compare them to \hessII{} results, given 
the absence of calibration between the two instruments. 
In this respect, the uncertainty on the \fermi-LAT absolute energy scale, $+2\%$/$-5\%$ \citep{2012ApJS..203....4A}, is to be considered as well. 

\section{Complementary tables}

\begin{table*}[h]
\caption{ Fit parameters to the \fermi-LAT light curve for P1 and P3 as a function of energy. An asymmetric Lorentzian and a log-normal function have been used, respectively.
The results for the estimated position of P1 using a Gaussian KDE ($\phi_{\rm P1}^{\rm KDE}$) are also shown. The errors on the latter estimates
were evaluated using a bootstrapping method. 
\label{tab:FermiP1P3FitResults}}             
\centering                          
\begin{scriptsize}  
\begin{tabular}{c | c c c c |c c }        
\hline\hline                        
Range & \multicolumn{4}{c|}{P1}   & \multicolumn{2}{c}{P3}    \\
\hline
             & $\phi_{\rm P1}$    & $\sigma_{\rm L}$ &  $\sigma_{\rm T}$ & $\phi_{\rm P1}^{\rm KDE}$ &  $\phi_{\rm P3}$ & $\sigma_{\rm P3}$   \\    
(GeV)        & (phase units)  & (phase units)    & (phase units)     & (phase units)         & (phase units)& (phase units)\\    
\hline                        
\hline
1-3   &  0.12979 $\pm$ 0.00014 & 0.008354 $\pm$ 0.000094  & 0.01462 $\pm$ 0.00032  & 0.1328 $\pm$0.0001 & 0.27454$\pm$  0.00058   & 0.3126 $\pm$ 0.0026\\   
 3-10  & 0.12964 $\pm$ 0.00029 & 0.006194 $\pm$ 0.00019   & 0.02372 $\pm$ 0.00078  & 0.1342 $\pm$0.0002 & 0.29985 $\pm$ 0.00077   & 0.2229 $\pm$ 0.0031\\    
 10-20 & 0.1298 $\pm$  0.0012 & 0.00364   $\pm$ 0.00080   & 0.0199 $\pm$ 0.0021    & 0.1355 $\pm$0.0006 & 0.3162 $\pm$ 0.0023     & 0.1567 $\pm$ 0.0065 \\    
 $>$20 & 0.1578 $\pm$  0.0026  & 0.0137   $\pm$ 0.0049    & 0.0001 $\pm$ 0.005     & 0.148  $\pm$0.004 & 0.4116 $\pm$  0.0043     & 0.387   $\pm$ 0.098 \\    
\hline                                   
\end{tabular}
\end{scriptsize}
\end{table*}

\begin{table*}[htb]
\begin{center}
\caption{List of $>80$ GeV photons in ascending phase value. Events are selected within a 
radius of $\theta_{\rm max}=0.8^\circ$ around the Vela pulsar from the 96 months \fermi-LAT data set. 
Columns give the phase, energy (in GeV), arrival time (Time, MJD), angular separation from the Vela pulsar ($\Delta\theta$), reconstructed angle with respect to the LAT boresight 
(Theta), angle between the reconstructed direction and the zenith line (Zenith, originates at the centre of Earth 
and passes through the spacecraft centre of mass), conversion type (0: FRONT; 1: BACK), and 
source probability evaluated using the \texttt{gtsrcprob} tool. This tool assigns to each photon the probabilities of originating from different sources.  
The source probabilities are shown for the Vela pulsar ($P_{\rm PSR}$) and the two sources that dominate at the highest energies: the Galactic diffuse emission ($P_{\rm GAL}$), 
and the Vela~X nebula ($P_{\rm VelaX}$). None of the events in the LW2 phase range display any peculiarity,
regarding  reconstructed angles in the instrument, conversion type, or zenith angle. 
\label{table:Fermi80GeVEvents}}
\begin{scriptsize}  
\begin{tabular}{ c c c | c c c c c c c}
\hline \hline
Phase & $E_\gamma$  &   Time   & $\Delta\theta$   & Theta        & Zenith    & Conv.      & $P_{\rm PSR}$    & $P_{\rm GAL}$       & $P_{\rm VelaX}$ \\
      & (GeV)       & (MJD)    & ($\degr$)        & ($\degr$)    & ($\degr$) &            &    (\%)          &  (\%)               &  (\%)     \\
\hline
0.006 &  99.6  & 54791.16     & 0.64 & 15.1  & 43.3 &  0     & -          & -  &-    \\ 
0.124 &  443.4 & 57170.69     & 0.78 & 15.5  & 64.4 &  0     & $10^{-3}$  & 55 & 43  \\
0.172 &  125.9 & 55198.06     & 0.75 & 22.4  & 65.1 &  0     & $8.10^{-4}$ & 86 & 1.3 \\
0.212 &  108.7 & 55116.93     & 0.57 & 30.3  & 36.3 &  0     & 1.0  &  47 & 45 \\
0.244 &  126.2 & 54751.05     & 0.42 & 21.3  & 25.3 &  1     & 6.0  &  49 & 36 \\
0.308 &  444.4 & 57483.43     & 0.60 & 51.7  & 87.4 &  1     & $7.10^{-2}$  &  57 & 22 \\
0.327 &  118.6 & 55785.65     & 0.53 & 24.6  & 66.9 &  1     & 3.4  &  39 & 41 \\
0.413 &  270.1 & 55528.44     & 0.62 & 42.9  & 88.0 &  1     & -    & -    & -\\
0.453 &  376.8 & 56667.56     & 0.16 & 30.0  & 23.1 &  0     & 99.3  &  0.4 & 0.3 \\
0.456 &  87.2  & 57534.34     & 0.14 & 18.1  & 50.8 &  0     & 96.8  &  1.0 & 2.0 \\
0.472 &  136.7 & 57109.13     & 0.36 & 38.5  & 64.0 &  1     & 76.6  &  11 & 9.0 \\
0.478 &  101.5 & 55530.03     & 0.34 & 41.0  & 84.1 &  1     & 78  &  11 & 10 \\
0.487 &  267.5 & 56483.62     & 0.74 & 21.2  & 71.0 &  0     & 11.5  &  43 & 37 \\
0.519 &  206.4 & 55154.10     & 0.33 & 35.8  & 81.4 &  1     &1.8  & 48  &39\\
0.564 &  93.7  & 56437.49     & 0.02 & 45.6  & 60.6 &  0     &99.8 & - & -\\
\hline \hline
\end{tabular} \\ 
\end{scriptsize}  
\end{center}
\end{table*}

\end{document}